\documentstyle[12pt,rotate]{article}

\input epsf.sty

\newcommand{\Sym}{\mathop{\rm S}}
\newcommand{\I}{{\cal I}}
\newcommand{\D}{{\cal D}}
\newcommand{\T}{{\cal T}}
\newcommand{\F}{{\cal F}}

\newcommand{\g}{{\sl g}}

\newcommand{\MS}{\overline{\rm MS}}

\begin{document}
\begin{titlepage}

\centerline{\large \bf Higher orders and infrared renormalon phenomenology}
\centerline{\large \bf in deeply virtual Compton scattering.}

\vspace{18mm}

\centerline{\bf A.V. Belitsky$^{a,b,}$\footnote{Alexander von
            Humboldt Fellow.}, A. Sch\"afer$^a$}

\vspace{18mm}

\centerline{\it ${^a}$Institut f\"ur Theoretische Physik, Universit\"at
                Regensburg}
\centerline{\it D-93040 Regensburg, Germany}
\centerline{\it ${^b}$Bogoliubov Laboratory of Theoretical Physics,
                Joint Institute for Nuclear Research}
\centerline{\it 141980, Dubna, Russia}

\vspace{25mm}

\centerline{\bf Abstract}

\hspace{0.8cm}

{\noindent
We present results for higher order perturbative corrections to Compton
scattering in the generalized Bjorken kinematics. The approach we have
used is based on the combination of two techniques: conformal operator
product expansion on the one side, and resummation of the fermion vacuum
insertions with consequent restoration of the full QCD $\beta$-function
via the naive nonabelianization assumption, on the other. These are terms
which are lost in the former approach. Due to the presence of the infrared
renormalon poles in the Borel transform of the resummed amplitude the
latter suffers from ambiguities which reflect the asymptotic character
of perturbation series. The residues of these IR renormalon poles give
an estimate for the size of power corrections in deeply virtual Compton
scattering.
}

\vspace{0.5cm}

\noindent Keywords: deeply virtual Compton scattering, conformal
operator product expansion, renormalons, higher twists

\vspace{0.5cm}

\noindent PACS numbers: 11.10.Hi, 11.30.Ly, 12.38.Bx, 12.38.Cy, 13.60.Fz

\end{titlepage}

%%%%%%%%%%%%%%%%%%%%%%%%%%%%%%%%%%%%%%%%%%%%%%%%%%%%%%%%%%%%%%%%%%%%%
\section{Introduction.}
%%%%%%%%%%%%%%%%%%%%%%%%%%%%%%%%%%%%%%%%%%%%%%%%%%%%%%%%%%%%%%%%%%%%%

A measurement of deeply virtual Compton scattering (DVCS)
\cite{Ji97,Rad96,Rad97,MRDGH} should yield important
information on the leading-twist non-forward parton
densities\footnote{The naming ``densities'' which is used by us
sometimes in the paper is not completely correct since it implies a
probabilistic interpretation for the corresponding entries, however, the
latter is lost for the non-forward functions. Therefore, it should
be understood only as a synonym.} which contain new information on the
strong interaction dynamics and could open a new window for the
exploration of the internal structure of the nucleon. To attain this
goal one should have theoretical control of higher orders of perturbation
theory as well as higher-twist corrections to the amplitudes. While QCD
perturbation theory is well established up to the multiloop (2,3,4-loop)
level, in practice all-order results for particular quantities are
inaccessible due to the essential complexity of many loop calculations.
The theory of power corrections is not settled completely. Even if
Operator Product Expansion (OPE) or its generalizations ---
factorization theorems\footnote{The proofs of factorization for DVCS and
diffractive meson production were given in Refs. \cite{Rad97,CFS96}.}
--- exist for a particular process such that the
power suppressed contributions can be expressed in terms of some
multiparton correlators, the latter cannot be evaluated quantitatively
mainly due to lack of a feasible non-perturbative approach. Therefore,
in fact already a rough order of magnitude estimate for the power
corrections would be extremely valuable. Furthermore the planned
experiments on DVCS will obtain primarily data at rather limited $Q^2$
(this is especially true for CEBAF experiments). Thus there is
legitimate concern that large higher-twist corrections might completely
obscure the interpretation of such experiments at least for some
kinematic regions.

Recently, it has been realized that both of the above issues can be
addressed by the study of the perturbative corrections in the large-$N_f$
limit of QCD \cite{renormalons,Braun94} to the leading twist
contributions. First of all, these calculations can be performed
exactly due to the relatively easy algebra. Second, when supplemented by
some assumptions, to be discussed below, it gives reasonable results when
compared to exact perturbative quantities available. Last but not the
least, by studying the ambiguities of the QCD perturbation series
one can get some insight into the size of power suppressed corrections
by the simple reason that only the sum of large orders of perturbation
theory and higher twist contributions is free from ambiguities and thus
physically relevant.

One remark should be added to all that has been said above. Since the
resummation of the vacuum insertions roughly corresponds to taking
into account only effects related to the running of the coupling,
all other radiative corrections are discarded at the same time. This
turns out sometimes to be an unreliable approximation. On the other
hand, by limiting oneself to the case when the $\beta$-functions is
zero, the so called hypothetical conformal limit, one has the advantage
of conformal covariance\footnote{Of course, even for $\beta = 0$
conformal invariance is broken by the renormalization of the field
operators, however, one can redefine the scale dimensions of the fields
\protect\cite{FGG73} and embed them into the original conformal
representation so that conformal covariance is preserved.} of
the theory and can make use of conformal operator product expansion
that allows to get strong restrictions on the form of the off-forward
part of the massless amplitudes in higher orders of QCD perturbation
theory. Therefore, below, when studying the higher order perturbative
corrections we will combine both approaches, i.e., renormalon chain
resummation and conformal OPE, which will amplify each other.

The main objects of the present investigation are structure functions
similar to those measured in deep inelastic scattering (DIS) but
generalized to non-forward kinematics. They appear as the coefficients
in the decomposition of the correlation function of two electromagnetic
currents into independent Lorentz tensor structures which individually
respect gauge invariance and are free of fictitious kinematical
singularities
\begin{eqnarray}
T_{\mu\nu} ( \omega, \zeta, Q^2 )
&=&
i \int d^4 z e^{iqz}
\langle h' |
T \left\{ J_\mu (0) J_\nu (z) \right\}
| h \rangle \nonumber\\
&=&
\left( - g_{\mu\nu} + \frac{q_\mu Q_\nu}{Qq} \right)
F^V ( \omega, \zeta, Q^2 )
+ \left( p_\mu - \frac{Qp}{Qq}q_\mu \right)
\left( p_\nu - \frac{qp}{Qq}Q_\nu \right)
\widetilde F^V ( \omega, \zeta, Q^2 ) \nonumber\\
&+& \frac{i}{Qp} \epsilon_{\mu \nu q p}
F^A ( \omega, \zeta, Q^2 ).
\end{eqnarray}
The form factors $F^{\mit\Gamma} ( \omega, \zeta )$ introduced above
differ only by an overall constant from the ones considered by us in
Ref. \cite{BelMul97}, namely $F^{\mit\Gamma} = -
\frac{1}{2}\F^{\mit\Gamma}$. The latter can be written in leading twist
approximation as a convolution of the non-perturbative
non-forward distributions and the perturbatively calculable
coefficient functions
\begin{equation}
\label{DVCS}
\F^{\mit\Gamma} ( \omega, \zeta ) = \int d x
\sum_{i=Q,G} {^i T^{\mit\Gamma}}( \omega, x, \zeta, Q^2 | \alpha_s )
{^i {\cal O}^{\mit\Gamma}} ( x, \zeta ).
\end{equation}
The former are defined as light-cone Fourier transformations of
non-local string operators sandwiched between appropriate hadronic
states ${^i{\cal O}^{\mit\Gamma}} (\lambda, \mu)$ \cite{Rad97}:
\begin{eqnarray}
\label{definition}
{^i{\cal O}^{\mit\Gamma}} (\lambda, \mu)
&=& \langle h' | \phi_i^* (\mu n) {\mit\Gamma}
\Phi \left[ \mu n, \lambda n \right]
\phi_i (\lambda n) |h \rangle \\
&=& \int d x e^{ i \mu (x - \zeta) - i \lambda x }\
\left\{
{^i{\cal O}^{\mit\Gamma}} (x, \zeta) \theta (x) \theta ( 1 - x)
-
{^{\bar i}{\cal O}^{\mit\Gamma}} (\zeta - x, \zeta)
\theta (\zeta - x) \theta (x + 1 - \zeta)
\right\}.\nonumber
\end{eqnarray}
The limits of integration on the right hand side of
Eq.~(\ref{definition}) can be deduced by studying the support properties
of the function  introduced above with Jaffe's approach \cite{Jaffe83}
(see also \cite{DG98} for a recent discussion) which results\footnote{
In the original derivation the author of Ref. \cite{Rad97} has used
the perturbative analyticity approach for studying support properties
of (multi)parton distributions \cite{Rad83} (see also \cite{EFP83}).}
in $- 1 + \zeta \leq x \leq 1$ \cite{Rad97}. Here $i = Q,G$ runs over
the parton species, ${\mit\Gamma}$ corresponds to different Dirac or
Lorentz structures, depending on the spin of the constituents involved
and $\Phi$ is a path ordered exponential. The parton momentum fractions
$x$ and $x-\zeta$ are the Fourier conjugated variables of the
light-cone positions $\lambda$ and $\mu$. The parameter $\zeta$ is
the skewedness of the distribution defined as a $+$-component of the
$t$-channel momentum. The perturbative expansion of the coefficient
function is given by
\begin{eqnarray*}
{^i T}^{\mit\Gamma}( \omega, x, \zeta, Q^2 | \alpha_s )
= {^i T_{(0)}} ( \omega, x ) +
{^i T_{(1)}^{\mit\Gamma}}( \omega, x, \zeta, Q^2 | \alpha_s )
+ {\cal O} (\alpha_s^2),
\end{eqnarray*}
with the leading order (LO) hand-bag contribution\footnote{In the
following consideration we will repeatedly omit the superscript $Q$ in
the LO amplitude.} ${^Q T_{(0)}} ( \omega, x ) = \omega / ( x \omega - 1 )
\pm ( x \to \zeta - x )$ with ``$+$''-sign corresponding to unpolarized
scattering and ``$-$'' to the spin-dependent case (and ${^G T_{(0)}} (
\omega, x ) = 0$). We have followed above the conventions introduced
by Radyushkin for DVCS \cite{Rad97}, namely,
$\omega = - 2(pQ)/Q^2$, and $\Delta_+ = q_+ - Q_+ = \zeta$, with $Q$
($p$) and $q$ ($p'$) being the incoming and outgoing momenta of the
photon (proton). They are related to those used by us in Ref.
\cite{Bel97}, (which are closely connected to the variables adopted
by the authors \cite{Ji97,MRDGH}) according to $\bar \omega \equiv
- ( \bar P \bar Q) /\bar Q^2 = \omega(2 - \zeta)/(2 - \omega \zeta)$,
$\eta = \zeta/(2 - \zeta)$, $t = (2x - \zeta)/(2 - \zeta)$, where the
averaged momenta are introduced as follows $\bar P = p + p'$, $\bar
Q = \frac{1}{2}( Q + q )$ and $\Delta_+ = \eta \bar P_+$. An advantage
of the first conventions is that the variable $x$ acquires a
simple partonic interpretation as the momentum fraction of the
incoming constituent, while the variable $t$ cannot be interpreted in
this way. However, its range does not depend on the longitudinal
asymmetry parameter $\eta$, contrary to $x$.

Considerable theoretical efforts have been undertaken recently to
explore the properties of DVCS in leading order: The evolution
kernels which govern logarithmic scaling violation have been evaluated
in Refs. \cite{BGR87,BB89,Ji97,Rad96,Rad97,FFGS97,kern,BR97,BelMul97}.
The solution of the renormalization group equation for the non-forward
distributions given above was found in \cite{Rad97,Bel97}, while
explicit numerical studies were performed in \cite{FFGS97,Bel97,Mank97}.
However, the values of $Q^2$ for which the hand bag approximation can be
trusted and gives satisfactory results need not be the same as in for
usual forward DIS \cite{DGPR97}. Therefore, this forces us to study the
effect of higher twists and large order perturbative corrections to the
amplitudes.

The first step for any analysis beyond LO is the evaluation of the NLO
coefficient functions. This issue was addressed first in \cite{JiOs97}
for the spin averaged singlet channel. In a previous paper \cite{BelMul97}
we evaluated all of them (recently these results were confirmed, see
Ref. \cite{Mank97NLO}) by taking advantage of conformal OPE for the
process in question. The way they were obtained will be discussed in
great detail in the next section.

Below we outline the ideas of the second approach which allows to restore
the effect of non-vanishing $\beta$-function and repeat some assumptions
which form the basis of this method. Its main idea is that the radiative
corrections related to the evolution of the coupling constant represent
the main source of large effects. In abelian theory these can be determined
by resummation of any number of fermion vacuum polarization insertions
in the gluon lines. In QCD this is no longer true, since there is a
number of other diagrams which contribute to the evolution of the
coupling. However, this statement can be justified in the limit of
an infinite number of fermion flavours $N_f \to \infty$. For moderate
$N_f$ it is clear that fermion bubbles do not produce sizable effects.
Therefore, in a second step, which has no strong theoretical foundation,
we add the sub-leading corrections in $N_f$ and restore the full QCD
$\beta$-function by hand from the corresponding $N_f$-dependent
coefficient \cite{BG95}:
\begin{equation}
\beta_0 = \frac{2}{3}N_f \to \frac{2}{3} N_f - \frac{11}{3}N_c,
\end{equation}
which is the first term in the perturbative expansion of the QCD
$\beta$-function
\begin{equation}
\beta_\epsilon = \mu \frac{\partial \g}{\partial \mu}
= - \epsilon \g + \beta (\g), \quad\mbox{with}\quad
\frac{\beta (\g)}{\g} = \frac{\alpha_s}{4 \pi} \beta_0 +
\left( \frac{\alpha_s}{4 \pi} \right)^2 \beta_1 + \dots .
\end{equation}

Thus as a first approximation one can restrict oneself to the mere
summation of the vacuum bubble chains substituted into the NLO
Feynman graphs. As we have noted at the very beginning in  the
large-$N_f$ limit the Borel transformed perturbation series for the
coefficient function $\T (\tau) = T_n/n! (-\tau/\beta_0)^n$ can be
obtained in closed form. From this the $n$-th order Wilson coefficient
$T_n$ can be obtained by taking the $n$-th derivative ($T_n =
1/(-\beta_0)^n d^n/d\tau^n |_{\tau=0} \T (\tau)$). This technique
gives good results for the quantities which are dominated by
renormalons, although this fact can be traced only a posteriori by
comparing the quantity in question with its exact value derived by
some other technique. The error $\Delta \T$ due to the asymptotic
character of the perturbation series for Wilson coefficients can
also be estimated from this Borel transformed series, i.e:
\begin{equation}
\Delta \T = \pm \frac{1}{\pi}
{\rm Im} \int_0^\infty d \tau
\exp\left(\frac{4\pi}{\beta_0 \alpha_s} \tau \right)
\T (\tau) .
\end{equation}
In the non-singlet case and in $d=4$ dimensions
$\T (\tau)$ results in a pole structure
\begin{equation}
\T (\tau) = \frac{1}{(Q^2)^\tau}
\left\{
\frac{R_0(\tau)}{\tau}
+ \frac{R_1(\tau)}{1 - \tau}
+ \frac{R_2(\tau)}{2 - \tau}
\right\},
\end{equation}
corresponding to power corrections of the form $1/Q^2,1/Q^4$ resulting
from the $1/(1 - \tau)$ and $1/(2 - \tau)$ infrared renormalon poles,
respectively. Hence $R_1(1)/Q^2$ and $R_2(2)/Q^4$ yield the
renormalon contributions to the off forward structure functions
$\F^{\mit\Gamma}$. Of course, they are completely unphysical and
should cancel against UV renormalon ambiguities of corresponding
twist-4 and twist-6 entries due to the mixing with low-twist
operators. But if one assumes that the ``genuine'' non-perturbative
contribution (which is essentially related to the specific hadron)
to the latter is small it may be used as a first approximation to
the magnitude of the non-leading twist corrections. This hypothesis is
referred to as ``ultraviolet dominance''. In some sense this is
the opposite assumption as the one made in the QCD sum rules approach
where one takes only the phenomenological condensates into account
which are accepted to be saturated by non-perturbative phenomena. Power
divergences due to renormalons are completely disregarded. Thus at
finite $\tau$ we can probe the non-perturbative effects seen as IR
renormalons. This approach to power suppressed contributions cannot
be claimed to be rigorous since it does not become exact in any limit
of QCD, but rather as a sophisticated guess checked in a number of
cases. In principle, the $1/N_f$ approximation kills the asymptotic freedom
of QCD and thus is inadequate to describe the real strong interaction
dynamics. However, it is by now established empirically that it allows
to define the position of renormalons provided one replaces the QED
$\beta$-function by real $\beta_{\rm QCD}$ (it gives reliable
prediction for the power $\tau$ in $1/Q^\tau$ corrections). Of course,
this consideration does not introduce any new information for the
processes where OPE is well established and the $1/Q^\tau$-behaviour
is known. But although OPE predictions being available for
about two decades for DIS they have been rarely used in practice due
to the absence of any systematic approaches for the study of
non-perturbative higher-dimensional operators. In the absence of
any adequate approach to higher twist phenomena\footnote{It is well
known that due to the higher dimensionality of the quantities involved
even the state-of-the-art QCD sum rules cannot claim an accuracy better
than $50\%-100\%$, although sometimes people quote smaller errors what
is of course not legitimate. Moreover, only few lowest moments can be
studied. Lattice QCD cannot help in this case for the time being due to
e.g. unsolved questions of operator mixing on the lattice.} it is
worthwhile to take advantage of the renormalon-based method and try to
fix the dependence of the power corrections on the momentum fraction
variables up to some numbers that fix the overall normalization. When
naively applied for various QCD observables the absolute magnitudes
$R(1)$ were not unreasonable \cite{renormalons}. All of this suggests
that it makes sense to apply the renormalon estimate to our problem.
We do know at least that for $\zeta\to 0$ the imaginary part of our
amplitude is related to the usual structure functions for which the
renormalon analyses gave sensible results \cite{IR_DIS97}.

According to the hypothesis of UV-dominance the shape and magnitude
of the higher twist corrections is determined by the intrinsic
ambiguity of the summation of the perturbative series and looks
like
\begin{equation}
\F^{\mit\Gamma} ( \omega, \zeta ) = \int d x
\left\{
T^{\mit\Gamma}( \omega, x, \zeta, Q^2 | \alpha_s )
+ \theta_1 \frac{\Lambda^2}{Q^2}
\Delta^{\mit\Gamma}_{\rm tw-4}(x, \omega, \zeta)
+ \theta_2 \frac{\Lambda^4}{Q^4}
\Delta^{\mit\Gamma}_{\rm tw-6}(x, \omega, \zeta)
\right\}
{\cal O}^{\mit\Gamma} ( x, \zeta ),
\end{equation}
where $\Lambda^2 = \mu^2 e^C$ (see below) and $\theta_i$ are adjustable
parameters which have to be fitted to data. Of course, the power
dependence in $Q^2$ is defined modulo logarithms of the hard scale $Q$
which are governed by the renormalization group equation. Unfortunately,
for multiparticle operators the evolution equations are extremely
complicated since the rang of the anomalous dimension matrix grows with
the moment of the correlator involved due to an increase of the number of
local operators mixed by renormalization. Therefore, the exact
evolution equation are of the Faddeev-type rather then of DGLAP form.
The solution of the problem at least for the non-singlet twist-3 sector
can be found by considering the integrated quantities which depend only
on one argument, and going to the multicolour limit. In the present case
the situation is complicated by the exclusiveness of the kinematics which
results in a mixing of the operators with total derivatives so that even
on leading twist level the anomalous dimensions turn out to be matrices
not numbers (to say nothing about higher twists).

In the present paper we present, apart from results on higher order
corrections derived from the application of conformal OPE, the
theoretical predictions for all-orders coefficient functions within the
NNA approximation and a model for the momentum fraction dependence of
the higher twist corrections coming from the assumption of the UV
dominance of the non-leading twist matrix elements. A complete and
thorough numerical analysis will be presented elsewhere together
with the implication of different models for the non-forward
distribution functions. Since the relative contributions of the
non-perturbative and perturbative corrections we have calculated will
depend on the functional forms assumed so will the final results.
For the time being the status of such models is still very
unsatisfactory.

%%%%%%%%%%%%%%%%%%%%%%%%%%%%%%%%%%%%%%%%%%%%%%%%%%%%%%%%%%%%%%%%%%%%%
\section{Beyond leading order.}
%%%%%%%%%%%%%%%%%%%%%%%%%%%%%%%%%%%%%%%%%%%%%%%%%%%%%%%%%%%%%%%%%%%%%
\label{NLO-CF}
It is well known since for a long time that the conformal
invariance of the theory puts severe restrictions on the possible
form of the amplitudes. In massless field theory this is true only at
tree level, but fails when the interaction is switched on due to
renormalization effects (we discard for the moment the effect of gauge
fixing in gauge theories since it is not of relevance for the
physical sector). The latter can be divided in two classes: i)
renormalization of the field operators and ii) running of the coupling
constant. While the first one is not too dangerous since after the
redefinition of the conformal representations by shifting the scale
dimensions of the fields, given originally in terms of the canonical
dimensions, by the anomalous ones, the theory respect conformal
covariance. However, the second effect inevitably breaks conformal
symmetry. Therefore, supposing the existence of nontrivial zero
$\g^*$ of the $\beta$-function ($\beta (\g^*) = 0$) conformaly
covariant OPE can be proven to exist even for the interacting
theory. Below we will shortly outline some of the points which
are of relevance for our further discussion.

%%%%%%%%%%%%%%%%%%%%%%%%%%%%%%%%%%%%%%%%%%%%%%%%%%%%%%%%%%%%%%%%%%%%%
\subsection{Conformal OPE and non-forward processes.}
%%%%%%%%%%%%%%%%%%%%%%%%%%%%%%%%%%%%%%%%%%%%%%%%%%%%%%%%%%%%%%%%%%%%%

There exists a complete basis of twist-2 conformal operators
${\cal O}$, which are labeled by the conformal spin $j$ and the scale
dimension $d_j = d_1 + d_2 + j$, constructed from products of the fields
of dimension $d_i$ and spin $s_i$ ($\nu_i = d_i + s_i - \frac{1}{2}$):
\begin{equation}
{\cal O}_{\mu_1 \dots \mu_l;j}
=
\Sym_{\mu_1 \dots \mu_n}
i\partial_{\mu_{j + 1}} \dots i\partial_{\mu_l}
\phi_{d_1}
\sum_{n = 0}^{j} C_n (\nu_1,\nu_2 )
i\partial_{\mu_1}
\dots
i\partial_{\mu_{j - n}}
i\stackrel{\leftrightarrow}{\partial}_{\mu_{j - n + 1}}
\dots
i\stackrel{\leftrightarrow}{\partial}_{\mu_j}
\phi_{d_2} - {\rm traces},
\end{equation}
with $\partial = \stackrel{\rightarrow}{\partial}
+ \stackrel{\leftarrow}{\partial}$ and
$\stackrel{\leftrightarrow}{\partial}
= \stackrel{\rightarrow}{\partial} - \stackrel{\leftarrow}{\partial}$
and coefficients $C_n (\nu_1,\nu_2 )$. Contracting this expression with
light-like vectors $n_\mu$ (its dual $n^* = p$ is a null vector\footnote{In
what follows the plus and minus indices in the place of the Lorentz
indices refer to a convolution with the vectors $n$ and $n^*$,
respectively.} along the opposite tangent to the light cone, defined
such that $p^2 = 0$, $np = 0$) (${\cal O}_{jl} = {\cal O}_{++ \dots +;j}$)
we obtain immediately the well know expression for the conformal
operators
\begin{equation}
{\cal O}_{jl}
= (i\partial_+)^l
\phi_{d_1}
P_j^{\nu_1 - \frac{1}{2},\nu_2 - \frac{1}{2}}
\left(\stackrel{\leftrightarrow}{\partial}_+ / \partial_+ \right)
\phi_{d_2},
\end{equation}
where $P$ are the usual Jacobi polynomials \cite{BE53_2,PBM}.

From dimensional counting alone we can easily write for the $T$-product
of two local (scalar) currents $J_A$ and $J_B$
\begin{equation}
\label{OPE}
T \left\{ J_A (0) J_B (x) \right\}
= \sum_{j = 0}^{\infty}
\left(
\frac{1}{x^2}
\right)^{\frac{1}{2}(d_A + d_B - d_j + j)}
\sum_{l = j}^{\infty} C_{jl} \ x_{\mu_1} \dots \mu_{\mu_l}
{\cal O}_{\mu_1 \dots \mu_l;j} (0).
\end{equation}
Since the operators ${\cal O}$ transform covariantly with respect to
the algebra of the conformal group there exists a relation between the
coefficients. This relation can be found by applying, for instance,
the generator of the special conformal transformation to both sides
of Eq. (\ref{OPE}). This leads to a recurrence relation with the
following solution:
\begin{equation}
C_{jl} = (-1)^{l-j}
\frac{\Gamma ( d_j + j ) \Gamma
\left( \frac{1}{2}( d_B - d_A + d_j - j + 2l ) \right)}{
\Gamma ( l - j + 1 ) \Gamma ( d_j + l )
\Gamma \left( \frac{1}{2}( d_B - d_A + d_j - j ) \right) } C_{jj}.
\end{equation}
Performing the sum over $l$ we get finally
\begin{eqnarray}
&&T \left\{ J_A (0) J_B (x) \right\} \\
&&\qquad = \sum_{j = 0}^{\infty} C_{jj}
\left(
\frac{1}{x^2}
\right)^{\frac{1}{2}(d_A + d_B - d_j + j)}
x_{\mu_1} \dots x_{\mu_j}
{_1F_1}
\left(
\left.
{ \frac{1}{2}(d_B - d_A + d_j + j)
\atop d_j + j }
\right| x_\sigma \partial_\sigma \right)
{\cal O}_{\mu_1 \dots \mu_j;j}(0). \nonumber
\end{eqnarray}
Thus the advantage of the conformal OPE for the non-forward processes
is that the Wilson coefficients are fixed entirely by symmetry up
to $C_{jj}$-coefficients which can be fixed from the forward matrix
elements known, for instance, from DIS.

For the case of two electromagnetic currents in order to avoid
complications which arise due to the fact that the latter carry Lorentz
indices it is enough to consider the trace and the antisymmetic part
of the amplitude. In this way the off-diagonal part (in the conformal
basis) of the amplitude is fixed unambiguously. (Remark: The difference
between the non-diagonal analogues ($F_V$ and $\widetilde F_V$) of the
forward structure functions $F_1$ and $F_2$ comes entirely from the
forward Wilson coefficient function of DIS since as long as the former
carry the same operator content the corrections to eigenfunctions are
the same in both cases). Thus, we have to leading twist accuracy
\begin{eqnarray}
\label{COPE-EMcurrents}
&&{\cal P}^{\mit\Gamma}_{\mu\nu}\
J_\mu (0) J_\nu (x) \\
&&\qquad = \sum_{j = 0}^{\infty}
C_{jj}^{\mit\Gamma}
\left(
\frac{1}{x^2}
\right)^{\frac{1}{2}(2d_J - d_j + j + 1)}
x_{\mu_0} x_{\mu_1} \dots x_{\mu_j}
{_1F_1}
\left(
\left.
{ \frac{1}{2}( d_j + j + 1 )
\atop d_j + j + 1 }
\right| x_\sigma \partial_\sigma \right)
{\cal O}^{\mit\Gamma}_{\mu_0\mu_1 \dots \mu_j;j}(0), \nonumber
\end{eqnarray}
where ${\mit\Gamma}$ labels the polarization of the amplitude in
question, and ${_1F_1}$ is a confluent hypergeometric function
\cite{BE53_1,PBM}. The projectors are defined as follows
${\cal P}^V_{\mu\nu} = g_{\mu\nu}$ for the spin averaged case and
${\cal P}^{A}_{\mu\nu} = i \epsilon_{\mu\nu +-}$ for the spin
dependent one.

In the interacting theory the conformal operators will mix under
renormalization in MS-type schemes so that covariance will
be lost. Since the mixing matrix is triangular due to Lorentz
invariance (only the operators with the same Lorentz spin $l$ can
mix with each other) its eigenvalues are given by the diagonal
matrix elements which coincide with the anomalous dimensions known
for forward scattering ${\gamma^{\mit\Gamma}_j}$. Thus, the
renormalization group equation for the latter must be diagonalized
first. This can be done by a finite transformation expressing
conformal operators ${\cal O}_{jl}$ of the free theory through the
multiplicatively renormalizable ones denoted by $\widetilde{\cal O}_{jl}$:
\begin{equation}
{\cal O}_{jl} =
\sum_{k = 0}^{j} B_{jk}\widetilde{\cal O}_{jl},
\end{equation}
with the transformation matrix $\hat B$. It is expressed in
terms of the off-diagonal $\hat\gamma^{\rm ND}$ matrix elements of
the anomalous dimension matrix $\hat\gamma$ of the conformal operators
${\cal O}_{jk}$. On the other hand $\hat\gamma^{\rm ND}$ can be fixed
entirely from the constraints coming from the algebra of dilatation
and conformal generators, namely \cite{Mue94,Mue97} $\left[ \hat a
(l,\frac{3}{2}) + \hat\gamma^c(l), \hat\gamma \right] = 0$ for $\beta
= 0$ (all entries will be specified below). As was shown in Ref.
\cite{Mue97} this holds true in every order of perturbation theory
for $\beta = 0$. Therefore, in the interacting theory conformal
OPE is of the same form as Eq. (\ref{COPE-EMcurrents}) provided we
replace the canonical scale dimensions $d_j^{\rm can}$ of the
operators by $d_j = d_j^{\rm can} + 2{^i\gamma^{\mit\Gamma}_j}$ and
rotate the tree level conformal operators to the covariant ones.

Taking into account all that has been said above we sandwich
Eq. (\ref{COPE-EMcurrents}) between hadronic states and get
after some simple algebra the following result for the Fourier
transformed amplitude
\begin{equation}
\label{COPE}
\F^{\mit\Gamma} ( \omega, \zeta ) =
\sum_{i} \sum_{j = 0}^{\infty}
{^iC_j^{\mit\Gamma}} (\alpha_s,Q^2/\mu^2) \
\omega^{j + 1}
{_2F_1} \left( \left.
{
1 + j + {^i\gamma^{\mit\Gamma}_j},\
2 + j + {^i\gamma^{\mit\Gamma}_j}
\atop
4 + 2j + 2\, {^i\gamma^{\mit\Gamma}_j}
}
\right| \omega \zeta \right)
\langle h' | {^i \widetilde {\cal O}}^{\mit\Gamma}_{jj} | h \rangle ,
\end{equation}
where ${_2F_1}$ is the hypergeometric function \cite{BE53_1,PBM}, and
${^iC_j^{\mit\Gamma}}$ is the Wilson coefficient for forward
scattering.

Since the product of the coefficient function and the eigenfunctions
of the evolution kernels is a scheme independent quantity we can use
any scheme we want. However, for other schemes the manifestly conformaly
covariant form of the OPE (\ref{COPE}) will be hidden.

%%%%%%%%%%%%%%%%%%%%%%%%%%%%%%%%%%%%%%%%%%%%%%%%%%%%%%%%%%%%%%%%%%%%%
\subsection{NLO corrections.}
%%%%%%%%%%%%%%%%%%%%%%%%%%%%%%%%%%%%%%%%%%%%%%%%%%%%%%%%%%%%%%%%%%%%%

The main idea of the conformal approach for the evaluation of the
coefficient functions consists in the combination of the information
coming from conformal OPE and from ordinary factorization theorems.
On the one hand we have the prediction for the amplitude from conformal
OPE outlined in the previous section with an input gained from the forward
scattering coefficient functions and anomalous dimensions. On the other
hand NLO factorization tells us that the total $\alpha_s$-correction is
generated by two sources: i) $\alpha_s$-corrections to the coefficient
function and ii) $\alpha_s^2$-corrections to the evolution kernels which
lead to a modification of the eigenfunctions of the evolution equation
of order ${\cal O} (\alpha_s)$. Thus, since the latter can be
fixed from conformal constraints we can get information about the
former from the combined use of the first and second representation
of the amplitude. Schematically, it reads
\begin{equation}
{^i T} =
{^i T^{\rm COPE}}
-
{^i T^{\rm kernel}},
\end{equation}
where the conventions we have used are self-explanative. In this
section we will only deal with the $s$-channel contribution. The
crossed amplitude can be obtained at the end by a mere substitution,
namely $x \to \zeta - x$.

The general conformal decomposition of the non-forward distributions
looks like
\begin{equation}
{^i{\cal O}} (x, \zeta)
= \sum_{k = Q,G} \sum_{j=0}^{\infty}
{^{ik}\phi}_j (x, \zeta | \alpha_s)
\langle h' | {^k{\cal O}}_{jj} | h \rangle ,
\end{equation}
where the partial conformal waves are generalized, beyond one-loop
level, to non-polynomial functions which are the subject of the
constraints.

The correction to an eigenfunction is defined completely in terms of
the $\hat B$-matrix
\begin{equation}
\label{eigenfunctionsER-BL}
\phi_j = \sum_{k = j}^{\infty}
\frac{x \bar x }{N_k (\frac{3}{2})}
C^{\frac{3}{2}}_k (2x - 1)
B_{kj}, \qquad
N_j (\nu) = 2^{ - 4 \nu + 1 }
\frac{ \Gamma^2 (\frac{1}{2}) \Gamma ( 2 \nu + j )}
{( \nu + j ) \Gamma^2 (\nu) \Gamma (j + 1)},
\end{equation}
which in the conformal limit is defined completely in terms of
the special conformal symmetry breaking matrix $\hat\gamma^c$
\cite{Mue94,BelMul97} via the following relation
\begin{equation}
\label{B-matrix}
\hat B = \frac{1}{1 + {\cal J} \hat\gamma^c},
\quad\mbox{with}\quad
{\cal J} \hat M = \frac{M_{jk}}{a(j,k,3/2)} \theta_{j,k+1},
\quad\mbox{for}\quad \forall \, \hat M-{\rm matrix},
\end{equation}
where $a(j,k,\nu) = 2 [ (j + 1)(j + 2) - (k + 1)(k + 2)
+ (2\nu - 3)(j - k) ]$. To obtain $\phi_j$ in the simplest way one
should derive the differential equation for the latter by using the
eigenvalue equation for the Gegenbauer polynomials as well as the
identity $a(j,k,3/2)B_{jk} = - \{\hat\gamma^c \hat B\}_{jk}$ which
follows from the definition of the $\hat B$-matrix (\ref{B-matrix}).

For the singlet quark case, which is relevant for our present analysis,
they have the form\footnote{The subtleties related to the spectral
restriction $0 \leq y \leq \zeta$ will be clarified in section
\ref{NNA-kernel}.}
\begin{equation}
{^{Qk}\phi}_j (x, \zeta | \alpha_s)
= \left\{
\delta_{Qk} \delta (x - y)
+ \frac{\alpha_s}{2\pi} {^{Qk}{\mit\Phi}} (x, y, \zeta)
\right\}
\otimes
\frac{1}{N_j (\nu) \zeta^{j + 1}}
\left[
\frac{y}{\zeta} \left( 1 - \frac{y}{\zeta} \right)
\right]^{\nu - \frac{1}{2}}
C_j^{\nu} \left( 2 \frac{y}{\zeta} - 1 \right),
\end{equation}
where $\nu = \nu (k)$, $k$ sums up the parton species ($Q,G$), and we
introduced the shorthand notation $\otimes \equiv \int dy$.

To fix the normalization of the amplitude coming from conformal OPE
we insert the conformal partial waves expansion into the LO result
(\ref{DVCS}) which followed from factorization and get as prediction for
the DVCS function:
\begin{equation}
\F^{\mit\Gamma}_{LO} ( \omega, \zeta )
= - 2
\sum_{j = 0}^{\infty}
B( j + 1 , j + 2 )
\omega^{j + 1}
{_2F_1} \left( \left.
{
1 + j ,\
2 + j
\atop
4 + 2j
}
\right| \omega \zeta \right)
\langle h' | {^Q{\cal O}^{\mit\Gamma}_{jj}} | h \rangle .
\end{equation}
Here the conformal matrix elements of the non-forward distributions
are defined in terms of the conformal operators sandwiched between
the physical states in question
\begin{equation}
\left\{\!\!\!
\begin{array}{c}
{^Q\!{\cal O}^V} \\
{^Q\!{\cal O}^A}
\end{array}
\!\!\!\right\}_{jl}
\!=
\bar{\psi} (i \partial_+)^l\!
\left\{\!\!\!
\begin{array}{c}
\gamma_+ \\
\gamma_+ \gamma_5
\end{array}
\!\!\!\right\}
\!C^{\frac{3}{2}}_j\!
\left( \frac{\stackrel{\leftrightarrow}{D}_+}{\partial_+} \right)
\!\psi .
\end{equation}
The gluonic case differs by the replacements $\frac{3}{2} \to \frac{5}{2}$
and $(j,l) \to (j,l) - 1$, and ${\mit\Gamma}$ stands now for the Lorentz
rather than Dirac structures:
\begin{equation}
\left\{\!\!\!
\begin{array}{c}
{^G\!{\cal O}^V} \\
{^G\!{\cal O}^A}
\end{array}
\!\!\!\right\}_{jl}
\!=
G_{+ \mu} (i \partial_+)^{l-1}\!
\left\{\!\!\!
\begin{array}{c}
g_{\mu\nu} \\
i\epsilon_{\mu\nu-+}
\end{array}
\!\!\!\right\}
\!C^{\frac{5}{2}}_{j - 1}\!
\left(
\frac{\stackrel{\leftrightarrow}{D}_+}{\partial_+}
\right)
\!G_{\nu +}.
\end{equation}

%%%%%%%%%%%%%%%%%%%%%%%%%%%%%%%%%%%%%%%%%%%%%%%%%%%%%%%%%%%%%%%%%%%%%
\subsubsection{Polarized sector: gluons.}
%%%%%%%%%%%%%%%%%%%%%%%%%%%%%%%%%%%%%%%%%%%%%%%%%%%%%%%%%%%%%%%%%%%%%
\label{Polarized-Gluon}

%%%%%%%%%%%%%%%%%%%%%%%%%%%%%%%%%%%%%%%%%%%%%%%%%%%%%%%%%%%%%%%%%%%%%
%            Figure 1
%%%%%%%%%%%%%%%%%%%%%%%%%%%%%%%%%%%%%%%%%%%%%%%%%%%%%%%%%%%%%%%%%%%%%

\begin{figure}[t]
\begin{center}
\vspace{3.8cm}
\hspace{-1cm}
\mbox{
\begin{picture}(0,220)(270,0)
\put(0,-30)                    {
\epsffile{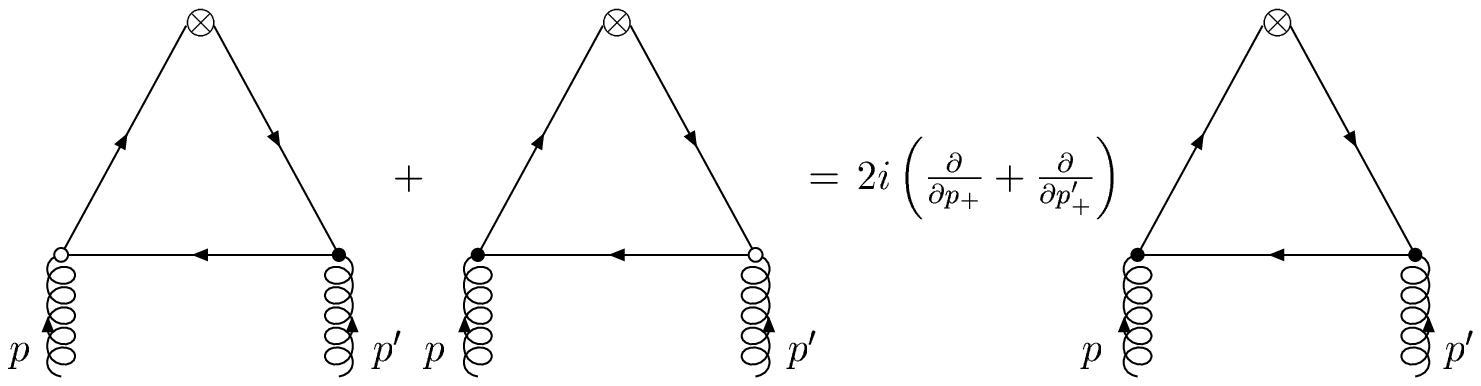}
                               }
\end{picture}
}
\end{center}
\vspace{-8cm}
\caption{\label{anomaly}
One-loop Feynman diagrams for the $QG$ special conformal anomaly matrix.
The empty vertex corresponds to the modified Feynman rules according to
Eq. (\protect\ref{modified-vertex}) while the full point denotes the usual
quark-gluon vertex. The blob with cross stands for the quark conformal
operator which mixes with the gluonic one due to renormalization.}
\end{figure}

As we have seen above the correction to the eigenfunction
${^{QG}{\mit\Phi}}$ can be found with the help of the special
conformal anomaly matrix in the quark-gluon channel. For the
$QG$-sector the derivation of the conformal Ward identities can
be performed in abelian gauge theory following the same reasoning
as for the non-singlet $QQ$-sector. This is true since in the
conformal limit, the symmetry breaking parts of the kernels do not
contain the Casimir operator $C_A$ of the adjoint representation of
$SU(3)$. Moreover, for non-vanishing $\beta$ the correct results can
be reconstructed by substituting the QCD $\beta$-function for the QED
one and, therefore, this treatment is sufficient to derive reliable
NLO predictions in QCD. The special conformal anomaly matrix in LO
looks like
\begin{equation}
{^{QG}\hat\gamma^c}
= \frac{\alpha_s}{2 \pi}\,
\bigl[
\, {^{QG}\hat\gamma}, \hat b (l)
\bigr]_-
+ {^{QG}\!\hat Z^\star_{[1]}},
\end{equation}
with $b_{jk} (l) = \theta_{jk} \left\{  2 (l + k + 3) \delta_{jk}
- [1 + (-1)^{j - k}] (2k + 3) \right\}$ and ${^{QG}\hat\gamma}$ being
the quark-gluon mixing anomalous dimension matrix. Thus, the only
difference compared to the analysis of Ref. \cite{Mue94} is that now we
have to evaluate the mixing renormalization matrix ${^{QG}\!\hat Z^\star}
= \frac{1}{\epsilon}{^{QG}\!\hat Z^\star_{[1]}} + \dots $ of the
conformal operator and the operator insertion
\begin{equation}
\label{modified-vertex}
\Delta^\g_- =
\g \frac{\partial}{\partial\g}
\int d^dx 2x_- {\cal L} (x),
\end{equation}
coming form the variation of the action with respect to the special
conformal transformation $\delta_-^C S = - \frac{\beta_\epsilon}{\g}
[\Delta^\g_-] + \dots$. This is the only source of symmetry breaking,
in contrast to the situation in the $QQ$-channel where subtleties
arise due to the quark equation of motion. Thus, only the
renormalization problem of the above mentioned operator has to be
solved\footnote{Square brackets mean minimally (MS) subtracted operators
\protect\cite{CollinsBOOK}.}
\begin{equation}
i [{^Q\!{\cal O}}_{jl}][\Delta^\g_-]
= i[{^Q\!{\cal O}}_{jl} \Delta^\g_-]
+ i\sum_{k = 1}^{j}
\left\{ {^{QQ}\!\hat Z^\star} \right\}_{jk}
[{^Q\!{\cal O}}_{k l - 1}]
+ i\sum_{k = 1}^{j}
\left\{ {^{QG}\!\hat Z^\star} \right\}_{jk}
[{^G\!{\cal O}}_{k l - 1}]
+ {\rm GVC},
\end{equation}
where GVC stands for the gauge-variant counterterms \cite{CollinsBOOK},
which are not of relevance for the present discussion since they cannot
affect the gauge invariant quantities we are interested in. To lowest
order in the coupling constant we have to calculate the diagram in Fig.
\ref{anomaly} with one of the vertices being replaced by the operator
insertion $i [\Delta^\g_-]$. To ${\cal O}(\alpha_s)$-accuracy we are
limited to, it reduces to $i [\Delta^\g_-]= i \g \mu^\epsilon \int dx 2
x_- \bar\psi{\not \!\! A} \psi$ and thus in the Feynman rules it results
in a mere differentiation with respect to the external gluon momenta of th
e last graph in Fig. \ref{anomaly} but with the familiar quark-gluon
vertices. Since the differentiation of the Gegenbauer polynomial in
momentum space is proportional to the $\hat b$-matrix acting on the
latter
\begin{equation}
\left(
\frac{\partial}{\partial p_+} + \frac{\partial}{\partial p'_+}
\right)
(p_+ + p'_+)^l C^\nu_j
\left(
\frac{p_+ - p'_+}{p_+ + p'_+}
\right)
=
\sum_{k = 0}^{j} b_{jk} (l)
(p_+ + p'_+)^{l - 1} C^\nu_j
\left(
\frac{p_+ - p'_+}{p_+ + p'_+}
\right),
\end{equation}
we finally get
\begin{equation}
{^{QG}\!\hat Z^\star_{[1]}}
= - \frac{\alpha_s}{2 \pi} \,
{^{QG}\hat\gamma} \, \hat b \, (l).
\end{equation}
Thus, the special conformal anomaly can be obtained immediately and
reads:
\begin{equation}
\label{QGCAM}
{^{QG}\hat\gamma^c}
= - \frac{\alpha_s}{2 \pi} \,
\hat{b}\, (l) \ {^{QG}\hat\gamma}.
\end{equation}

Following the discussion preceding this section we find from
Eq.\ (\ref{QGCAM}) that the corrections to the eigenfunctions are
completely expressed in terms of the shift operator:
\begin{equation}
\label{Phi}
{^{QG}{\mit\Phi}} (x, y, \zeta)
= ( \I - \D ) S (x, z) \otimes {^{QG}\!K^A} ( z, y, \zeta ),
\end{equation}
where the generalized evolution kernel reads\footnote{In this and
subsequent formulae we followed our previous convention and introduced
the generalized step functions \cite{BM97}: $\Theta^{m}_{i_1 i_2 ... i_n}
(x_1,x_2,...,x_n)=\int_{-\infty}^{\infty}\frac{d\alpha}{2\pi i}
\alpha^m \prod_{k=1}^{n}\left(\alpha x_k -1 +i0 \right)^{-i_k}$.}
\begin{equation}
\label{QG-A-kernel}
{^{QG}\!K^A} (x , x ', \zeta)
= 2 N_f T_F \Theta_{112}^1 ( x , x - \zeta, x - x ' ),
\end{equation}
and $S$ generates the shift of the Gegenbauer polynomials index:
\begin{equation}
S(x, y) \otimes [y \bar y]^{\nu - \frac{1}{2}}
C^\nu_j (2y - 1)
= \left. \frac{d}{d\rho} \right|_{\rho = 0}
[x \bar x]^{\nu - \frac{1}{2} + \rho} C^{\nu + \rho}_j (2x - 1).
\end{equation}
In Eq.\ (\ref{Phi}) $\I$ is an identity operator and
$\D$ extracts the diagonal part of any test function
$\tau (x, y)$ in its expansion with respect to a basis of
Gegenbauer polynomials $C^\nu_j$, i.e.
\begin{eqnarray*}
\int_{0}^{1} dx C^\nu_j (2x - 1)
\left\{
\begin{array}{c}
\I \\
\D
\end{array}
\right\}
\tau (x, y)
=
\sum_{ k = 0 }^{j}
\tau_{jk}
\left\{
\begin{array}{c}
1 \\
\delta_{jk}
\end{array}
\right\}
C^\nu_k (2y - 1).
\end{eqnarray*}
The diagonal matrix elements of the shift operator $S$ which will be
used below are for instance
\begin{equation}
S_{jj}
= 3\psi (2 + j) - 2\psi (4 + 2j) - \psi (1) - \frac{2 + 2j}{2 + j}.
\end{equation}
Here and below $\psi^{(n)} (j) = \frac{d^{n+1}}{d j^{n+1}} \ln \Gamma (j)$
denotes polygamma functions. The evolution kernel introduced in Eq.
(\ref{QG-A-kernel}) can be diagonalized with the following identity
\begin{equation}
\label{QGmixing}
\zeta
\int dy C_{j}^{\frac{3}{2}} \left( 2 \frac{y}{\zeta} - 1 \right)
{^{QG}\!K^A} ( y, x, \zeta )
= {^{QG}\gamma^A_j}
C_{j - 1}^{\frac{5}{2}}
\left( 2 \frac{x}{\zeta} - 1 \right) \ \mbox{with} \
{^{QG}\gamma^A_j} = - \frac{12 N_f T_F}{(j + 1)(j + 2)}.
\end{equation}
Technically, to get this equality one has to multiply both sides by
the factor $(y \bar y)^2$, differentiate three times with respect to
$y$ and use the relation
\begin{equation}
\left[ (y \bar y)^2 C^{5/2}_{j - 1} (2y - 1) \right]'''
= \frac{\Gamma (j + 4)}{6\Gamma (j)} C^{3/2}_{j} (2y - 1).
\end{equation}
The coefficient ${^{QG}\gamma^A_j}$ in front of the Gegenbauer
polynomial coincides with the DGLAP moments of the kernel
${^{QG}\!K^A} ( y, x, 0 )$ up to the common factor $6/j$, which
arises as a result of the conventional definition of the Gegenbauer
polynomials, namely, the coefficient of $x^j$ in $C_{j}^{\frac{3}{2}}
(x)$ is $3/j$ times the coefficient of $x^{j - 1}$ in
$C_{j - 1}^{\frac{5}{2}} (x)$; an additional factor of 2 comes
from the argument of the polynomial.

Since now we know the $\alpha_s$-correction to the eigenfunctions
of the two-loop $QG$-evolution kernel we can find coefficient
function in the corresponding channel. On the one hand conformal OPE
(\ref{COPE}) to NLO gives the following prediction
\begin{eqnarray}
\label{NLO-COPE}
{^G\!\F}^A_{\rm NLO}
&=& - 2 \frac{\alpha_s}{2\pi}
\sum_{j = 0}^{\infty} B (j + 1, j + 2) \omega^{j + 1}
\biggl\{
{^G\! C}_j^A
{_2F_1} \left( \left.
{
1 + j ,\
2 + j
\atop
4 + 2j
}
\right| \omega \zeta \right) \\
&&\hspace{3cm}+ {^{QG}\gamma^A_j}
\left. \frac{d}{d\rho} \right|_{\rho = 0}
{_2F_1} \left( \left.
{
1 + j + \rho,\
2 + j + \rho
\atop
4 + 2j + 2 \rho
}
\right| \omega \zeta \right)
\biggr\}
\langle h' | {^G{\cal O}^A_{jj}} | h \rangle , \nonumber
\end{eqnarray}
with the forward coefficient function \cite{AR88}
\begin{equation}
{^GC^A}_j (Q^2/\mu^2)
= 2 N_f T_f \frac{6}{j}
\frac{j}{(j + 1)(j + 2)}
\left\{
\ln \frac{- Q^2}{\mu^2}
- \psi (j + 1) + \psi (1) - 1
\right\}
\end{equation}

On the other hand factorization theorems tell us that the total
$\alpha_s$-correction to the amplitude reads
\begin{equation}
\label{NLO-factorization}
{^G\!\F}^A_{\rm NLO}
= \frac{\alpha_s}{2 \pi}
{^Q T_{(0)}} ( \omega, y ) \otimes
{^{QG}{\mit\Phi}} (x, y, \zeta)
\otimes {^G {\cal O}^A} ( x, \zeta )
+
{^G T_{(1)}^A}( \omega, x, \zeta, Q^2 | \alpha_s ).
\end{equation}
where ${^G T_{(1)}^A}$ is the quantity in question. To proceed further
we equate the right hand sides of Eqs. (\ref{NLO-COPE}) and
(\ref{NLO-factorization}) and use the formulae
\begin{eqnarray}
\label{Eq1}
&&\hspace{-1cm}T_{(0)} ( \omega, y ) \otimes S ( y, z )
\otimes
{^{QG}\!K^A} ( z, x, \zeta )
\otimes
{^G {\cal O}^A} ( x, \zeta )
=
- 2 \sum_{j = 0}^{\infty}
{^{QG}\gamma^A_j}
B( j + 1 , j + 2 ) \omega^{j + 1} \\
&&\qquad\times
\biggl\{
S_{jj} \,
{_2F_1} \left( \left.
{
1 + j ,\
2 + j
\atop
4 + 2j
}
\right| \omega \zeta \right)
+
\left. \frac{d}{d\rho} \right|_{\rho = 0}
{_2F_1} \left( \left.
{
1 + j ,\
2 + j + \rho
\atop
4 + 2j + 2 \rho
}
\right| \omega \zeta \right)
\biggr\}
\langle h' | {^G{\cal O}^A_{jj}} | h \rangle ,\nonumber\\
\mbox{and}\nonumber\\
\label{Eq2}
&&\hspace{-1cm}T_{(0)} ( \omega, y ) \otimes \ln (1 - \omega y)
{^{QG}\!K^A} ( y, x, \zeta )
\otimes
{^G {\cal O}^A} ( x, \zeta )
= - 2 \sum_{j = 0}^{\infty}
{^{QG}\gamma^A_j}
B( j + 1 , j + 2 ) \omega^{j + 1} \\
&&\qquad\times
\biggl\{
L_{jj} \,
{_2F_1} \left( \left.
{
1 + j ,\
2 + j
\atop
4 + 2j
}
\right| \omega \zeta \right)
-
\left. \frac{d}{d\rho} \right|_{\rho = 0}
{_2F_1} \left( \left.
{
1 + j + \rho ,\
2 + j
\atop
4 + 2j
}
\right| \omega \zeta \right)
\biggr\}
\langle h' | {^G{\cal O}^A_{jj}} | h \rangle \nonumber.
\end{eqnarray}
Here it is easy to recognize the first terms in the curly brackets
in Eqs. (\ref{Eq1}) and (\ref{Eq2}) as the diagonal parts of the
shift operator and the logarithm ${L}_{jj} = - [ \psi (j + 1)
- \psi (1) ]$, respectively. Extracting only the non-diagonal part
from these identities, and adding the equation which comes from the
forward coefficient function:
\begin{eqnarray}
\label{diagonal}
&& 2 \sum_{j = 0}^{\infty}
{^GC^A_j} (Q^2/\mu^2)
B( j + 1 , j + 2 ) \omega^{j + 1}
{_2F_1} \left( \left.
{
1 + j ,\
2 + j
\atop
4 + 2j
}
\right| \omega \zeta \right)
\langle h' | {^G{\cal O}^A}_{jj} | h \rangle \\
&&\qquad=
T_{(0)} ( \omega, z ) \otimes \left\{
{^{QG}\!K^A} (z , x , \zeta) \ln \frac{- Q^2}{\mu^2}
+ \left[ \D \ln (1 - z \omega) - 1 \right]
{^{QG}\!K^A} (z , x , \zeta) \right\}
\otimes {^G {\cal O}^A} ( x, \zeta ), \nonumber
\end{eqnarray}
we can restore the form of the amplitude coming from OPE. Subtracting
from the OPE result the correction to the eigenfunction coming from the
usual factorization approach we get the coefficient function we are
interested in
\begin{equation}
{^G T_{(1)}^A}( \omega, x, \zeta, Q^2 ) =
- \frac{\alpha_s}{2\pi}
T_{(0)} ( \omega, y ) \otimes
\biggl\{
{^{QG}\!K^A} ( y, x, \zeta )\, \ln \frac{- Q^2}{\mu^2}
+ \left[ \ln \left( 1 - y \omega \right) - 1 \right]
{^{QG}\!K^A} ( y, x, \zeta )
\biggr\}.
\end{equation}
By subtracting the logarithm of the hard scale we obtain the net
gluon coefficient function in the $\MS$ scheme.

%%%%%%%%%%%%%%%%%%%%%%%%%%%%%%%%%%%%%%%%%%%%%%%%%%%%%%%%%%%%%%%%%%%%%
\subsubsection{Polarized sector: quarks.}
%%%%%%%%%%%%%%%%%%%%%%%%%%%%%%%%%%%%%%%%%%%%%%%%%%%%%%%%%%%%%%%%%%%%%
\label{PolarizedSector}

In the quark sector there is no need to calculate the special
conformal anomaly matrix for extended kinematics since it is known
for the Efremov-Radyushkin-Brodsky-Lepage (ER-BL) case \cite{Mue94}.
It was shown in Ref. \cite{muel88} that the continuation to the whole
$x/\zeta,y/\zeta$-plane is unique. To leading order this results in
the substitution of the ordinary $\theta$-functions in the kernels
by the generalized ones \cite{BM97}, namely
\begin{equation}
k (x, y) \frac{\theta (x - y)}{1 - y}
\to
k \left( \frac{x}{\zeta}, \frac{y}{\zeta} \right)
\Theta^0_{11} (x - y , x - \zeta),
\end{equation}
with a crossed contribution obtained by the substitutions
$x \to \zeta - x$, $y \to \zeta - y$. The correspondence between
extended kernels and ER-BL ones is the following $K (x, y, \zeta = 1)
= - V_{\rm ER-BL} (x, y)$. This means that, in principle, one could
get the non-forward evolution kernels from their ER-BL analogues which
are known for over fifteen years rather then calculate them once more.
Coordinate space results could be reconstructed with a help of the
Fourier transformation \cite{BM97}.

Taking into account the effect of non-vanishing $\beta$-function
results in a modification of the eigenfunction of the NLO evolution
kernel by an additional term $\sim - \beta_0(\I - \D)S$, namely
\begin{equation}
\label{QQ-eigenfunction}
{^{QQ}{\mit\Phi}} (x, y, \zeta)
= ( \I - \D )
\left\{
S (x, z) \otimes
\left[
{^{QQ}\!K^A} ( z, y, \zeta )
- \frac{\beta_0}{2} \delta (z - y)
\right]
- {^{Q}\!G} ( y, x, \zeta )
\right\},
\end{equation}
where
\begin{eqnarray}
&&{^{QQ}\!K^A} (x , x ', \zeta) \\
&&\quad\qquad = C_F \,
\left[ \frac{x}{x - x'}
\Theta_{11}^0 (x , x - x ') + \frac{x - \zeta}{x - x'}
\Theta_{11}^0 (x - \zeta, x - x ')
+ \Theta_{111}^0 ( x , x - \zeta, x - x ' )
\right]_+ , \nonumber\\
&&{^{Q}\!G} ( x, x', \zeta ) \\
&&\quad\qquad = C_F \,
\left[
\frac{x' - \zeta}{x - x'}
\ln \left( \frac{x' - x}{x' - \zeta} \right)
\Theta_{11}^0 (x - \zeta, x - x ')
+
\frac{x'}{x - x'}
\ln \left( \frac{x' - x}{x'} \right)
\Theta_{11}^0 (x, x - x ')
\right]_+. \nonumber
\end{eqnarray}

From conformal OPE it follows that
\begin{eqnarray}
&& - 2 \sum_{j = 0}^{\infty}
{^QC^A_j} (Q^2/\mu^2)
B( j + 1 , j + 2 ) \omega^{j + 1}
{_2F_1} \left( \left.
{
1 + j ,\
2 + j
\atop
4 + 2j
}
\right| \omega \zeta \right)
\langle h' | {^Q{\cal O}^A_{jj}} | h \rangle \\
&&\qquad=
T_{(0)} ( \omega, z ) \otimes
\biggl\{
\delta (z - x)
-
\frac{\alpha_s}{2\pi}
\biggl[
{^{QQ}\!K^A} (z , x , \zeta) \ln \frac{- Q^2}{\mu^2}
+ \D \ln (1 - z \omega) {^{QQ}\!K^A} (z , x , \zeta) \nonumber\\
&&\hspace{5.55cm} - \frac{3}{2} {^{Q}\!K^b} ( z, x, \zeta )
- \D\, {^{Q}\!G} ( z, x, \zeta )
+ \frac{3}{2} \delta (z - x)
\biggr]
\biggr\}
\otimes {^Q {\cal O}^A} ( x, \zeta ) ,\nonumber
\end{eqnarray}
where the forward coefficient function is taken from Ref. \cite{ZVN94}:
\begin{eqnarray}
\label{Q-A-coeff-func}
{^Q C^A_j} (Q^2/\mu^2)
=
\biggl\{
1 + \frac{\alpha_s}{2\pi}C_F
\biggl[
2 S_{1,1} (j) \!\!\!&-&\!\!\! 2 S_2 (j)
+ S_1 (j)
\left(
\frac{3}{2}
+ \frac{1}{j + 1}
+ \frac{1}{j + 2}
\right) \\
&+&\!\!\! \frac{3}{j + 1}
- \frac{9}{2}
+ \left(
\frac{3}{2} - S_1 (j) - S_1 (j + 2)
\right) \ln \frac{- Q^2}{\mu^2}
\biggr]
\biggr\}. \nonumber
\end{eqnarray}
Here the additional kernel reads
\begin{eqnarray}
{^{Q}\!K^b} ( x, x', \zeta )
\!\!\!&=&\!\!\! C_F
\left[
\frac{x - \zeta}{x - x'}
\Theta_{11}^0 (x - \zeta, x - x ')
+
\frac{x}{x - x'}
\Theta_{11}^0 (x, x - x ')
\right]_+ .
\end{eqnarray}
In the derivation we have taken into account that the diagonal
matrix elements of the matrices introduced above read
\begin{eqnarray}
\frac{1}{C_F}
{^{Q}\!K^b}_{jj} &=& 2 [ \psi (j + 2) - \psi (1) ] - 2 ,\\
\frac{1}{C_F}
{^{Q}\!G}_{jj} &=&  \psi^{(1)} (j + 1) - \zeta (2)
- [ \psi (j + 1) - \psi (1) ]^2 , \\
\frac{1}{C_F}
{^{QQ}\!K^A}_{jj} & \equiv & \frac{1}{C_F} {^{QQ}\gamma^A_j}
=  \psi (j + 3) + \psi (j + 1) - 2 \psi (1) - \frac{3}{2} .
\end{eqnarray}
To recover the same form of the result as in Eq. (\ref{Q-A-coeff-func})
it is useful to modify these equalities by using the identity
$2 S_{1,1} (j) = S_1^2 (j) + S_2 (j)$ \cite{Yndurain79}, where $S_{l,m}
(j) = \sum_{k = 0}^{j} S_m (k) / k^l$, $S_m (k) = \sum_{i = 0}^{k} 1 /
i^m $, $\psi^{(1)} (j + 1) = \zeta (2) - S_2 (j)$ \cite{PBM}, and by
noting that with these conventions ${^{Q}\!G}_{jj} = - 2 C_F S_{1,1}(j)$.

Thus, the same steps as before give
\begin{eqnarray}
\label{unpolQCF}
{^Q T_{(1)}^A}( \omega, x, \zeta, Q^2 ) \!\!\!&=&\!\!\!
- \frac{\alpha_s}{2\pi}
T_{(0)} ( \omega, y ) \otimes
\biggl\{
{^{QQ}\!K^A} ( y, x, \zeta )\,
\ln \frac{- Q^2}{\mu^2}
- \frac{3}{2} {^{Q}\!K^b} ( y, x, \zeta )
+ \frac{3}{2} \delta (y - x) \nonumber \\
&+& \ln \left( 1 - y \omega \right)
{^{QQ}\!K^A} ( y, x, \zeta )
- {^{Q}\!G} ( y, x, \zeta )
\biggr\} .
\end{eqnarray}
In the limit $\zeta = 1$ we obtain from Eq. (\ref{unpolQCF}) the
results of Refs. \cite{Bra83} for the $\pi\to\gamma\gamma$ transition
form factor $F_{\pi\gamma\gamma}$.

%%%%%%%%%%%%%%%%%%%%%%%%%%%%%%%%%%%%%%%%%%%%%%%%%%%%%%%%%%%%%%%%%%%%%
\subsubsection{Unpolarized sector: quarks and gluons.}
%%%%%%%%%%%%%%%%%%%%%%%%%%%%%%%%%%%%%%%%%%%%%%%%%%%%%%%%%%%%%%%%%%%%%

For the quark vector channel the only additional term comes from the
forward DIS coefficient function \cite{ZVN94} and reads
$C_F/(j + 1)(j + 2)$. It is easy to check that this eigenvalue
corresponds to the evolution kernel $\Theta^0_{111}$, namely
\begin{equation}
\int dx C_j^{3/2} \left( 2\frac{x}{\zeta} - 1 \right)
\Theta^0_{111} (x, x - \zeta, x - y)
=
\frac{1}{(j + 1)(j + 2)}
C_j^{3/2} \left( 2\frac{y}{\zeta} - 1 \right).
\end{equation}

As we have noted in the previous subsection in order to verify this
and similar eigenvalue problems there is no need to perform the
integration rather one should differentiate $\nu + \frac{1}{2}$-times
both sides of the equations multiplied by $(y \bar y)^{\nu -
\frac{1}{2}}$ with respect to $y$ and use the eigenvalue equation for
the Gegenbauer polynomials of index $\nu$. To reduce the multi-argument
$\Theta$-functions to the two-argument ones it is enough to use the
following relations
\begin{eqnarray}
\Theta^n_{ijk} (x_1, x_2, x_3)
&=& \frac{1}{x_1 - x_2}
\left\{
\Theta^{n - 1}_{i-1jk} (x_1, x_2, x_3)
-
\Theta^{n - 1}_{ij-1k} (x_1, x_2, x_3)
\right\} \nonumber\\
&=& \frac{1}{x_1 - x_2}
\left\{
x_2 \Theta^n_{i-1jk} (x_1, x_2, x_3)
-
x_1 \Theta^n_{ij-1k} (x_1, x_2, x_3)
\right\}.
\end{eqnarray}
Thus, the problem to find the off-forward analogues to some forward
functions is reduced to the solution of the inverse problem of
determining a potential from the known eigenvalues
\begin{equation}
\int dx C_j^{\nu} \left( 2\frac{x}{\zeta} - 1 \right)
K (x, y, \zeta)
=
E_j(\nu)
C_j^{\nu} \left( 2\frac{y}{\zeta} - 1 \right),
\end{equation}
which in general turns out to be quite nontrivial. If this problem
would be solved in general we would be able to reconstruct the
diagonal part of the kernel in the basis of Gegenbauer polynomials
from the corresponding forward analogue and thus the whole NLO
corrections to the ER-BL-type evolution kernel and not only its
off-diagonal part.

Observing that
\begin{equation}
\Theta^0_{111} (x, x - \zeta, x - y)
=
{^{QQ}\!K^V} ( x, y, \zeta )
- {^{Q}\!K^b} ( x, y, \zeta )
- \frac{1}{2} \delta (x - y),
\end{equation}
and that the corrections to the $QQ$-eigenfunctions in the spin-dependent
and -averaged cases are the same we easily obtain\footnote{Note a misprint
in Eq. (40) of Ref. \protect\cite{BelMul97} where the sign in the
brackets of the fourth term should be changed.}:
\begin{eqnarray}
\label{Nonpolarized-quarks}
{^Q T_{(1)}^V}( \omega, x, \zeta, Q^2 ) \!\!\!&=&\!\!\!
- \frac{\alpha_s}{2\pi}
T_{(0)} ( \omega, y ) \otimes
\biggl\{
{^{QQ}\!K^V} ( y, x, \zeta )\,
\ln \frac{- Q^2}{\mu^2}
- \frac{5}{2} {^{Q}\!K^b} ( y, x, \zeta )
+ \delta (y - x) \nonumber \\
&+& \left[
\ln \left( 1 - y \omega \right) + 1
\right]
{^{QQ}\!K^V} ( y, x, \zeta )
- {^{Q}\!G} ( y, x, \zeta )
\biggr\} , \\
{^G T_{(1)}^V}( \omega, x, \zeta, Q^2 ) \!\!\!&=&\!\!\!
- \frac{\alpha_s}{2\pi}
T_{(0)} ( \omega, y ) \otimes
\biggl\{
{^{QG}\!K^V} ( y, x, \zeta )\,
\ln \frac{- Q^2}{\mu^2}
- \frac{1}{2}
\left[
{^{QG}\!K^V} ( y, x, \zeta ) + {^{QG}\!K^A} ( y, x, \zeta )
\right]
\nonumber \\
&+& \ln \left( 1 - y \omega \right) {^{QG}\!K^V} ( y, x, \zeta )
\biggr\} .
\end{eqnarray}
The calculation of the gluon coefficient function quoted here does not
introduce any new specific features and its derivation runs along the
same line as before. The momentum space evolution kernels
${^{Q(Q,G)}\!K^V}$ involved above read
\begin{eqnarray}
{^{QQ}\!K^V} (x , x ', \zeta) &=& {^{QQ}\!K^A} (x , x ', \zeta),\\
{^{QG}\!K^V} (x , x ', \zeta)
&=& {^{QG}\!K^A} (x , x ', \zeta)
- 4 N_f T_F \frac{( x - x' )}{x'( x' - \zeta )}
\Theta_{111}^0 ( x , x - \zeta, x - x ' ). \nonumber
\end{eqnarray}

%%%%%%%%%%%%%%%%%%%%%%%%%%%%%%%%%%%%%%%%%%%%%%%%%%%%%%%%%%%%%%%%%%%%%
\subsubsection{Remark.}
%%%%%%%%%%%%%%%%%%%%%%%%%%%%%%%%%%%%%%%%%%%%%%%%%%%%%%%%%%%%%%%%%%%%%

To get the corrections beyond NLO one should expand the amplitude
coming from conformal OPE (\ref{COPE}) up to the required $n$-th order.
However, these results would correspond to the special scheme where
the conformal covariance is preserved. To be able to make the
predictions in MS-type schemes we have to perform finite renormalization
with $\hat B$-matrix which has to be known to $n$-th order. The latter
requires evaluation of the special conformal anomaly $\gamma^c$ in
the corresponding channel. For instance, to NNLO this calculation goes
along the same line as developed in Ref. \cite{MikhsRad} for the
two-loop non-singlet exclusive ER-BL evolution kernel, provided we
replace the ordinary Feynman rules, by modified ones similar to those
used by us in section \ref{Polarized-Gluon}. (Technically this analysis
is of the same complexity as in the paper mentioned above). However,
even then, this prediction will be only valid for the special case of
vanishing $\beta$-function $\beta = 0$ which, of course, may turn out
to be far from reality. Therefore, it is worthwhile to study the
importance of conformal symmetry breaking effects introduced by the
running of the QCD coupling constant. This subject will be addressed
in the next section.

%%%%%%%%%%%%%%%%%%%%%%%%%%%%%%%%%%%%%%%%%%%%%%%%%%%%%%%%%%%%%%%%%%%%%
\section{Resummation of fermion vacuum insertions.}
%%%%%%%%%%%%%%%%%%%%%%%%%%%%%%%%%%%%%%%%%%%%%%%%%%%%%%%%%%%%%%%%%%%%%

In this section we will present results for the resummation of the
fermion bubble chains in the NLO coefficient functions of the DVCS
amplitude. The idea of the NNA approximation is based on the
observation that the effects related to the evolution of the coupling
constant can be represented as a source of potentially large
perturbative corrections. Its extraction can give important information
on the higher order perturbative contributions.

The resummation of the fermion loops in the NLO coefficient function
leads to a factorial growth of the series
\begin{equation}
T - T_{(0)} = \sum_{n = 0}^{\infty} d_n n! (-\alpha_s \beta_0)^n ,
\end{equation}
which convergence radius is zero. Therefore, due to the asymptotic
character of perturbation series the naive resummation is meaningless
and to get reasonable predictions we should truncate the series at the
order $n_0$ at which the ratio of two successive terms is of the order 1.
The first neglected term will give an estimate for the ambiguity of this
approximation. This ambiguity is known to be power suppressed. This last
point, discussed at length in the Introduction, will be used below for
the construction of a model for the non-leading twist corrections.

%%%%%%%%%%%%%%%%%%%%%%%%%%%%%%%%%%%%%%%%%%%%%%%%%%%%%%%%%%%%%%%%%%%%%
\subsection{Coefficient function.}
%%%%%%%%%%%%%%%%%%%%%%%%%%%%%%%%%%%%%%%%%%%%%%%%%%%%%%%%%%%%%%%%%%%%%
\label{NNA-CF}

%%%%%%%%%%%%%%%%%%%%%%%%%%%%%%%%%%%%%%%%%%%%%%%%%%%%%%%%%%%%%%%%%%%%%
%            Figure 2
%%%%%%%%%%%%%%%%%%%%%%%%%%%%%%%%%%%%%%%%%%%%%%%%%%%%%%%%%%%%%%%%%%%%%

\begin{figure}[t]
\begin{center}
\vspace{5.7cm}
\hspace{-1cm}
\mbox{
\begin{picture}(0,220)(270,0)
\put(0,-30)                    {
\epsffile{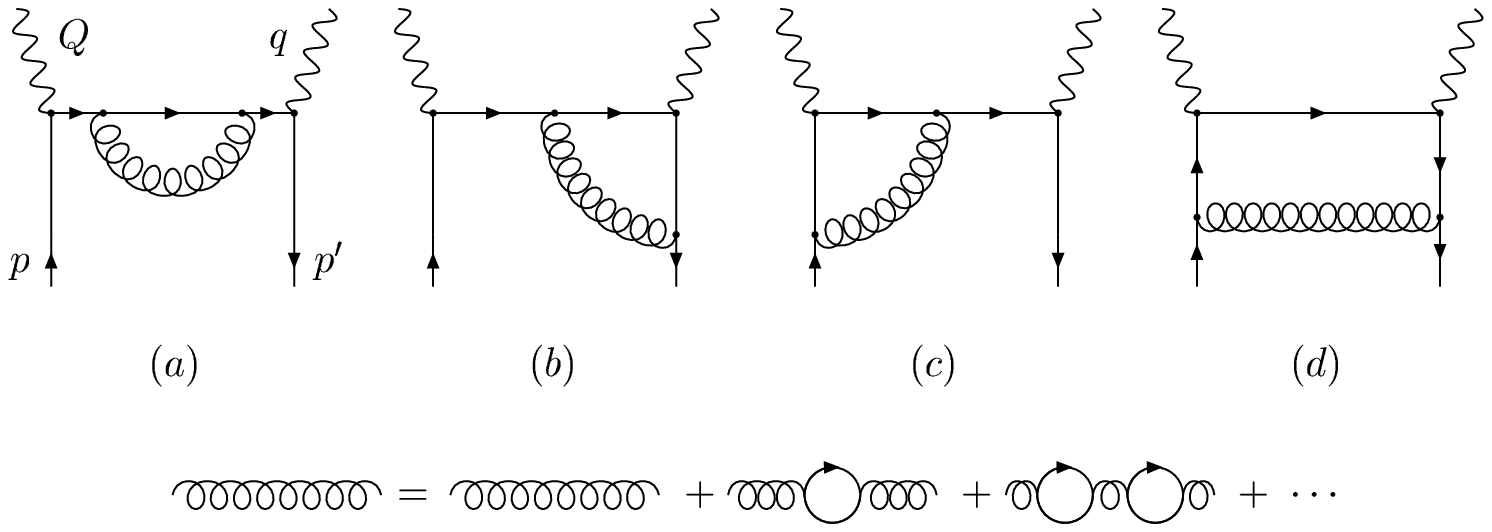}
                               }
\end{picture}
}
\end{center}
\vspace{-8cm}
\caption{\label{NLO-NNA} NLO Feynman diagrams ($s$-channel only) with
resummed fermion vacuum insertions.}
\end{figure}

In the following we adhere to the convention $b \equiv - \beta_0$
for the first coefficient of the $\beta$-function. After resummation
of the fermion vacuum polarization blobs (the sum is defined up to
infinity in the sense of a principal value (PV) prescription for
the poles in the Borel integral, although in practice the truncation
of the series at its minimal term is completely equivalent to this) in
the NLO coefficient function (see Fig. \ref{NLO-NNA}) and with appropriate
renormalization we get as result for the coefficient function for the
$\mit\Gamma$-amplitude (the details of this calculation can be found
in Appendix A)
\begin{equation}
\label{FermionBubbleAmplitude}
T^{\mit\Gamma} ( \omega, x, \zeta, Q^2 | \alpha_s )
= - \frac{2C_F}{b}\, {\rm PV}\!\!\int_0^\infty \frac{d \tau}{\tau}
e^{- 4 \pi / (\alpha_s b )\tau}
\left\{
\T^{\mit\Gamma} (0, \tau | \omega, x, \zeta, Q^2 )
-
\widetilde\T^{\mit\Gamma} (- \tau, 0 | \omega, x, \zeta, Q^2 )
\right\},
\end{equation}
where\footnote{In what follows we omit the dependence of the Borel
transform $\T^{\mit\Gamma} (\epsilon , \tau)$ on the momentum fractions
as well as other kinematical variables.} $\widetilde\T^{\mit\Gamma}
(\tau , 0)$ is a Borel transform of $\T^{\mit\Gamma}$ with respect
to the first argument. Note that the second term can be expressed in
terms of the integral of the function $\T^{\mit\Gamma} (\tau , 0)$
itself
\begin{equation}
\int_0^\infty \frac{d \tau}{\tau}
e^{- 4 \pi / ( \alpha_s b )\tau}
\widetilde\T^{\mit\Gamma} (- \tau, 0 )
=
\int_0^{-\frac{\alpha_s b}{4\pi}} \frac{d \tau}{\tau}
\T^{\mit\Gamma} (\tau, 0 ).
\end{equation}
A straightforward calculation leads to the following representation of
the Borel amplitude
\begin{equation}
\T^{\mit\Gamma} (\epsilon, \tau ) =
F (\epsilon , \tau )
D^{\mit\Gamma} (\epsilon , \tau ),
\end{equation}
with a common factor $F$
\begin{equation}
F (\epsilon , \tau )
=
\left( \frac{\mu^{*2}}{-Q^2} \right)^\tau
\left[ 6 \frac{ \Gamma ( 1 + \epsilon ) \Gamma^2 ( 2 - \epsilon )}{
\Gamma ( 4 - 2 \epsilon ) }
\right]^{\frac{\tau}{\epsilon} - 1}
\frac{\pi \tau}{\sin (\pi \tau)}
\frac{ \Gamma (2 - \epsilon)}{
\Gamma (1 - \epsilon + \tau) \Gamma (3 - \epsilon - \tau)}
\end{equation}
($\mu^{*2} = 4\pi \mu^2$) and a function $D$ which is specific for the
particular channel (${\mit\Gamma} = A,V$). As can be seen form Eq.
(\ref{FermionBubbleAmplitude}) in practice we need the function
$F$ only for vanishing first or second argument. We can easily get
\begin{equation}
F( 0, \tau ) =
\left( \frac{\mu^{2} e^C}{-Q^2} \right)^\tau
\frac{1}{(1 - \tau)(2 - \tau)}.
\end{equation}
where $C$ differs for different subtraction schemes:
$C_{\rm MS} = \frac{5}{3} + \ln 4\pi + \psi (1)$, $C_{\MS} =
\frac{5}{3}$.
\begin{equation}
F(\tau , 0 ) =
\frac{1}{6} \frac{(1 - \tau)}{( 2 - \tau )}
\frac{\Gamma (4 - 2 \tau) }{ \Gamma (1 + \tau) \Gamma^3 (2 - \tau) }.
\end{equation}

The contributions of individual diagrams to the function $D$ in the
polarized and unpolarized channels are given in Appendix A. Below we
just quote these expressions for those particular values of its arguments
which enter Eq. (\ref{FermionBubbleAmplitude}).

%%%%%%%%%%%%%%%%%%%%%%%%%%%%%%%%%%%%%%%%%%%%%%%%%%%%%%%%%%%%%%%%%%%%%
\subsubsection{Polarized case.}
%%%%%%%%%%%%%%%%%%%%%%%%%%%%%%%%%%%%%%%%%%%%%%%%%%%%%%%%%%%%%%%%%%%%%

\begin{eqnarray}
\label{D_A}
D^A(0, \tau) &=&
T_{(0)} ( \omega, x )
\biggl\{
\frac{3}{( 1 - x\omega )^\tau}
- \left[
\frac{2}{1 + \tau} - (2 - \tau) \frac{ 1 - x \omega}{\zeta \omega}
\right]
{_2F_1} \left.\left(
{
1 + \tau,\ 1 + \tau
\atop
2 + \tau
}
\right| x \omega \right) \nonumber\\
&&- \frac{1}{( 1 - \zeta \omega )^\tau}
\left[
\frac{2}{1 + \tau} + (2 - \tau) \frac{ 1 - x \omega}{\zeta \omega}
\right]
{_2F_1} \left.\left(
{
1 + \tau,\ 1 + \tau
\atop
2 + \tau
}
\right| \frac{( x - \zeta )\omega}{ 1 - \zeta \omega } \right)
\biggr\}, \\
D^A(\tau, 0) &=&
T_{(0)} ( \omega, x )
\biggl\{
\frac{3 - \tau^2}{ 1 - \tau}
-
\frac{2 - \tau}{1 - \tau}
\left[ 1 - (1 - \tau) \frac{ 1 - x \omega}{\zeta \omega}
\right]
{_2F_1} \left.\left(
{
1 ,\ 1 - \tau
\atop
2 - \tau
}
\right| x \omega \right) \nonumber\\
&&- \frac{2 - \tau}{1 - \tau}
\left[
1 + (1 - \tau) \frac{ 1 - x \omega}{\zeta \omega}
\right]
{_2F_1} \left.\left(
{
1 ,\ 1 - \tau
\atop
2 - \tau
}
\right| \frac{( x - \zeta )\omega}{ 1 - \zeta \omega } \right)
\biggr\}.
\end{eqnarray}

%%%%%%%%%%%%%%%%%%%%%%%%%%%%%%%%%%%%%%%%%%%%%%%%%%%%%%%%%%%%%%%%%%%%%
\subsubsection{Unpolarized case.}
%%%%%%%%%%%%%%%%%%%%%%%%%%%%%%%%%%%%%%%%%%%%%%%%%%%%%%%%%%%%%%%%%%%%%

\begin{eqnarray}
\label{D_V}
D^V(0, \tau) &=&
T_{(0)} ( \omega, x )
\biggl\{
\frac{3}{( 1 - x\omega )^\tau}
- \frac{1}{ 1 + \tau }
\left[
2 - (2 - \tau ( 1 - \tau ) ) \frac{ 1 - x \omega}{\zeta \omega}
\right]
{_2F_1} \left.\left(
{
1 + \tau,\ 1 + \tau
\atop
2 + \tau
}
\right| x \omega \right) \nonumber\\
&&- \frac{1}{ 1 + \tau }
\frac{1}{( 1 - \zeta \omega )^\tau}
\left[
2 + (2 - \tau ( 1 - \tau ) ) \frac{ 1 - x \omega}{\zeta \omega}
\right]
{_2F_1} \left.\left(
{
1 + \tau,\ 1 + \tau
\atop
2 + \tau
}
\right| \frac{( x - \zeta )\omega}{ 1 - \zeta \omega } \right)
\biggr\}, \\
D^V (\tau, 0) &=& D^A (\tau, 0).
\end{eqnarray}

The general feature manifested by the results we have just derived is
the existence of a few IR renormalon poles in the Borel plane. This is
a common point for all space-like processes \cite{IR_DIS97} which tells
us what kind of power suppressed contributions should be added to make
physical quantities free from ambiguities.

%%%%%%%%%%%%%%%%%%%%%%%%%%%%%%%%%%%%%%%%%%%%%%%%%%%%%%%%%%%%%%%%%%%%%
\subsection{Extended ER-BL evolution kernel.}
\label{NNA-kernel}
%%%%%%%%%%%%%%%%%%%%%%%%%%%%%%%%%%%%%%%%%%%%%%%%%%%%%%%%%%%%%%%%%%%%%

The question of the NNA-corrections to the eigenfunctions of the
generalized Efremov-Radyushkin-Brodsky-Lepage evolution equation
was studied by us in Ref. \cite{BelMul97} (see also \cite{GosKiv97}).
They are given by Eq. (\ref{eigenfunctionsER-BL}) but now the conformal
anomaly $\gamma^c$ in the $\hat B$-matrix should be shifted by the
term proportional to the $\beta$-function $\hat\gamma^c \to \hat\gamma^c
+ 2 \frac{\beta}{\g}\hat b$. This follows from the conformal constraints
extended to the case of a running coupling \cite{Mue94}. It was shown
there that the eigenfunctions of the extended ER-BL evolution kernels
with the fermion renormalon chains resummed to all orders leads to a
shift of the index $\frac{3}{2}$ of the Gegenbauer polynomial by the
amount $-\beta(\g)/\g$:
\begin{equation}
\left[
\frac{x}{\zeta} \left( 1 - \frac{x}{\zeta} \right)
\right]^{1 - \beta/\g }
C_j^{3/2 - \beta/\g }
\left( 2 \frac{x}{\zeta} - 1 \right).
\end{equation}
The NNA-anomalous dimension matrix is diagonalized in this basis
\begin{equation}
\hat B^{-1} {^{QQ}\!\hat K_{\rm NNA}} \hat B
= {^{QQ}\!\hat\gamma^{\rm NNA}},
\end{equation}
and its diagonal matrix elements coincide with the forward anomalous
dimension of the DGLAP evolution equation with fermion-loop insertions
resummed to all orders \cite{Gra97,Mankiewicz97,Mik97,GosKiv97}:
\begin{eqnarray}
{^{QQ}\!\gamma^{\rm NNA}_j}
\left( \tau = \frac{\alpha_s}{4 \pi}\beta_0 \right)
= C_F F(\tau, 0)
\Bigl\{
1 - \tau
&-& \frac{(1 - \tau)^2(2 - \tau)}{(1 - \tau + j)(2 - \tau + j)}
\nonumber\\
&+& (4 - 2 \tau)
\left[
\psi (2 - \tau + j) - \psi (2 - \tau)
\right]
\Bigr\}.
\end{eqnarray}
Since the LO evolution kernels are the same in the polarized and
unpolarized cases the eigenfunctions and eigenvalues are also the
same. But, unfortunately, the eigenfunctions obtained by resumming
only the fermion bubble chains with consequent restoration of the
full $\beta$-function has nothing to do with reality since
as has been observed in Ref. \cite{Mue95} there is significant
cancelation in the evolution kernels between different conformal
symmetry breaking parts for $N_f = {\rm light\ flavours}$, namely,
between the special conformal anomaly and the $\beta$-function
term. However, in the present approach while the latter is taken into
account the former is discarded completely. The breaking of the
NNA approximation for the anomalous dimension of forward deep
inelastic scattering has been observed also in Ref.
\cite{Mankiewicz97} which supports our conclusions.

The above result could be seen from the expressions for the
eigenfunction found in the previous section \ref{PolarizedSector}.
Adhering to the NNA we should neglect all terms in Eq.
(\ref{QQ-eigenfunction}) except for those $\sim \beta$ and to ${\cal O}
(\alpha_s)$ accuracy (for brevity we put $\zeta = 1$) we find
\begin{equation}
\left\{ I - (\I - \D) \tau S \right\} \otimes
[y \bar y] C_j^{3/2} (2 y - 1)
=
{\cal N}_j (\tau) [y \bar y]^{1 - \tau}
C_j^{3/2 - \tau} (2 y - 1) + {\cal O} (\tau^2) ,
\end{equation}
where $\tau = \frac{\alpha_s}{4 \pi}\beta_0$, and ${\cal N}_j (\tau)
= 1 + \tau S_{jj}$ comes from the diagonal part of the shift operator
which affects only the overall normalization. In this way we
support the hypothesis of NNA since in Eq. (\ref{QQ-eigenfunction})
$\beta_0$ stands for the full QCD $\beta$-function rather then only
for its fermionic piece up to limitation about its quantitative
validity.

Note that the argument of the Gegenbauer polynomial $C^\nu_j(t)$
is defined on the segment $-1 \leq t \leq 1$, thus the above
equalities are defined for a distribution with the support
$0 \leq x \leq \zeta$ \cite{Rad96}. However, if one exploits
the identity
\begin{eqnarray}
&&\Theta_{11}^0 (x, x - \zeta)
\left[
\frac{x}{\zeta}
\left( 1 - \frac{x}{\zeta} \right)
\right]^{\nu - 1/2}
C^\nu_j \left( 2 \frac{x}{\zeta} - 1 \right) \\
&&\hspace{5cm}= 2^{1 - 2\nu}
\frac{\Gamma \left( \frac{1}{2} \right) \Gamma (j + 2 \nu)}{
\Gamma (\nu) \Gamma (j + \nu + \frac{1}{2}) \Gamma (j + 1)}
\int_{0}^{1} dt (t \bar t)^{j + \nu - 1/2}
\delta^{(j)} (\zeta t - x), \nonumber
\end{eqnarray}
and can understand the latter in the sense of a mathematical
distribution, i.e. to get a meaningful result it should be
convoluted first with some smooth function before integration
over $t$ \cite{Rad96}, one can then restore the support properties
to the whole range $0 \leq x \leq 1$.

%%%%%%%%%%%%%%%%%%%%%%%%%%%%%%%%%%%%%%%%%%%%%%%%%%%%%%%%%%%%%%%%%%%%%
\section{Renormalon model for the higher twist corrections.}
%%%%%%%%%%%%%%%%%%%%%%%%%%%%%%%%%%%%%%%%%%%%%%%%%%%%%%%%%%%%%%%%%%%%%

The model for higher twist contributions can be traced from the
estimation of the ambiguity in the resummation of the perturbation
series and is given by the imaginary part of the Borel integral
which appears due to the necessity to deform the integration path
in the complex plane of the $\tau$-parameter to escape from singularities
(this leads to the undetermined overall sign owing to different
possibilities to close the integration contour around the pole)
\begin{equation}
- \frac{2C_F}{b}\frac{1}{\pi}\, {\rm Im}\! \int_0^\infty
\frac{d \tau}{\tau} e^{- 4 \pi / (\alpha_s b )\tau}
\T^{\mit\Gamma} (0, \tau | \omega, x, \zeta, Q^2 )
= \pm \sum_{{\rm tw}-n}
\left( \frac{\mu^{2} e^C}{-Q^2} \right)^{(n - 2)/2}
\Delta^{\mit\Gamma}_{{\rm tw}-n} (x, \omega, \zeta).
\end{equation}
There are only two poles in the Borel plane which are located in the
points $\tau = 1,2$. Thus, the ambiguities generate only two types of
power corrections corresponding to twist-4 and 6 operators. Of
course, one can generate even higher power correction by considering
the perturbative diagrams beyond one-loop order. However, in view of
the success of the leading IR renormalon model we will limit ourselves
to the one-loop approximation. From Eqs. (\ref{D_A}), (\ref{D_V}) of
the preceding section we get for the spin-dependent sector
\begin{eqnarray}
&&\Delta^A_{{\rm tw}-4}(x, \omega, \zeta) =
\frac{4C_F}{b}
\frac{T_{(0)} ( \omega, x )}{(x - \zeta)\omega}
\biggl\{
\frac{ (1 - \zeta \omega)(1 - x \omega + \zeta \omega)}{\zeta
\omega (x - \zeta) \omega} \ln (1 - \zeta \omega) \\
&&\qquad\qquad+
\frac{1}{x\omega}
\left[
1 - x\omega - \frac{(2x - \zeta)\omega (1 - \zeta \omega)}{x\omega
(x - \zeta)\omega}
\right]
\ln (1 - x\omega)
+ \frac{1}{1 - x\omega}
\left[
\frac{1}{2}(x - \zeta)\omega - \frac{1 - \zeta \omega}{x\omega}
\right]
\biggr\}, \nonumber\\
&&\Delta^A_{{\rm tw}-6}(x, \omega, \zeta) =
\frac{2C_F}{b}
T_{(0)} ( \omega, x )
\biggl\{
\frac{ 1 - \zeta \omega}{(x - \zeta)^3 \omega^3} \ln (1 - \zeta \omega)
-
\left[
\frac{1}{(x\omega)^3}
+ \frac{1 - \zeta \omega}{ (x - \zeta)^3\omega^3}
\right]
\ln (1 - x\omega) \nonumber\\
&&\qquad\qquad- \frac{1}{2(1 - x\omega)^2}
\biggl[
\frac{2}{(x\omega)^2}
- \frac{1 - x\omega}{x\omega}
\left(
3 - \frac{x\omega ( 1 - \zeta \omega )}{( x - \zeta )^2\omega^2}
\right) \\
&&\hspace{7.4cm}-
\frac{1 - \zeta \omega}{(x - \zeta)^2\omega^2}
\left(
( 1 - \zeta \omega )( 2x - \zeta )\omega
- \frac{ x\omega }{( x - \zeta )\omega}
\right)
\biggr]
\biggr\}, \nonumber
\end{eqnarray}
and for the spin-averaged one
\begin{eqnarray}
&&\Delta^V_{{\rm tw}-4}(x, \omega, \zeta) =
\Delta^A_{{\rm tw}-4}(x, \omega, \zeta), \nonumber\\
&&\Delta^V_{{\rm tw}-6}(x, \omega, \zeta) =
\frac{2C_F}{b}
T_{(0)} ( \omega, x )
\biggl\{
\left(
1 + 2 \frac{1 - x \omega}{\zeta \omega}
\right)
\frac{ 1 - \zeta \omega}{(x - \zeta)^3 \omega^3}
\ln (1 - \zeta \omega) \\
&&\qquad\qquad -
\left[
\frac{1}{(x\omega)^3}
\left(
1 - 2 \frac{1 - x \omega}{\zeta \omega}
\right)
+ \frac{1 - \zeta \omega}{ (x - \zeta)^3\omega^3}
\left(
1 + 2 \frac{1 - x \omega}{\zeta \omega}
\right)
\right]
\ln (1 - x\omega) \nonumber\\
&&\qquad\qquad- \frac{1}{2(1 - x\omega)^2}
\biggl[
\frac{2}{(x\omega)^2}
- \frac{1 - x\omega}{x\omega}
\left(
3 + 2 \frac{ 2 - 3 x \omega )}{ x \omega \zeta \omega}
\right) \\
&&\hspace{5.4cm}-
\frac{1 - \zeta \omega}{(x - \zeta)^2\omega^2}
\left(
1 + 2 \frac{1 - x \omega}{\zeta \omega}
\right)
\left(
1 - 3x \omega + \zeta \omega
+ \frac{ x\omega }{( x - \zeta )\omega}
\right)
\biggr]
\biggr\}. \nonumber
\end{eqnarray}
We should note that the twist-4 result can be considered as more
reliable since the only contribution to it comes from the IR
renormalons in the coefficient function for the twist-2 operators.
For the twist-6 part there is additional input from IR ambiguities
in the point $\tau = 1$ from the coefficient function of the twist-4
correlators. Thus the above presented results for tw-6 include only
part of the total contribution. However, since the latter is
suppressed by extra two powers of the momentum transfer its
particular shape is not of relevance for us and we have cited
them just for completeness.

There is another source of power suppressed contributions which
comes from the non-con\-ver\-gent behaviour of the perturbation theory
and is due to instantons \cite{Lipatov}. However, they lead to strongly
suppressed contributions due to high values of $\tau$ and thus
can safely be discarded.

\section{Summary and outlook.}

In this paper we have presented the theoretical framework for studying
higher order and higher twist corrections based on the combination
of two different formalisms. The first one is based on restrictions
for the amplitudes of the massless theory coming from the algebra of
the collinear conformal group in the hypothetical limit of vanishing
$\beta$-function. We restore the dependence on the latter with the
second approach which resums the fermion vacuum insertions to all
orders in the coupling. Taken alone this approximation is insufficient
since fermion loops do not dominate the radiative corrections to the
amplitudes. However, supplied with the idea of
NNA one may hope that they do and that thus all important perturbative
corrections are taken into account. The validity of this approximation
can be checked only by comparison with exact results. It turns out that
not all quantities are approximated with good quality by this methods,
but only those which are dominated by the renormalon poles. Due to the
latter the resummed amplitude is plagued by the uncertainties that
manifest the asymptotic character of the perturbative series. It was
established that they are power suppressed and thus mimic power
corrections to the amplitude. This tells us that both higher orders and
higher twist should be treated simultaneously to escape from
ambiguities intrinsic to the perturbative series. By accepting that
the non-perturbative higher dimensional operators are dominated
by their UV renormalon poles it is possible to construct rough model
for the momentum fraction dependence of the power suppressed
contributions.

In conclusion we have resummed fermion vacuum polarization bubbles in
the coefficient function of the DVCS amplitude. It is known that NNA
approximation overestimates radiative corrections. A more realistic
estimate is given by the semi-sum of the exact NLO (section
\ref{NLO-CF}) and NNA (section \ref{NNA-CF}) results. By using the UV
dominance hypothesis we give an estimate of the shape and magnitude of
the higher twist contributions. The numerical analysis will be performed
in a separate publication. For that we have to accept one of the models
for leading twist non-forward distribution. Several of them are already
on a market \cite{Radyushkin-model,JiMelSon97,Pol97}.

\vspace{0.5cm}

Note added: After the present study was finished we learned about
Refs. \cite{JiOs98,Collins98} where the leading twist factorization
for DVCS was proved to all orders of perturbation theory.

\vspace{0.5cm}

We wish to thank M. Maul for the collaboration at an early stage
of the work and A.V. Radyushkin for comment. A.B. was supported
by the Alexander von Humboldt Foundation and partially by Russian
Foundation for Fundamental Research, grant N 96-02-17631.

\vspace{0.5cm}

\appendix

\setcounter{section}{0}
\setcounter{equation}{0}
\renewcommand{\theequation}{\Alph{section}.\arabic{equation}}

%%%%%%%%%%%%%%%%%%%%%%%%%%%%%%%%%%%%%%%%%%%%%%%%%%%%%%%%%%%%%%%%%%%%%
\section{Fermion bubbles in the DVCS coefficient function.}
%%%%%%%%%%%%%%%%%%%%%%%%%%%%%%%%%%%%%%%%%%%%%%%%%%%%%%%%%%%%%%%%%%%%%
\label{DVCS-coeff-fuct-NNA}
%%%%%%%%%%%%%%%%%%%%%%%%%%%%%%%%%%%%%%%%%%%%%%%%%%%%%%%%%%%%%%%%%%%%%
\subsection{Spin-dependent scattering.}
%%%%%%%%%%%%%%%%%%%%%%%%%%%%%%%%%%%%%%%%%%%%%%%%%%%%%%%%%%%%%%%%%%%%%

In the treatment of the polarized sector we resolved the
$\gamma_5$-ambiguity following Braaten's recipe \cite{Bra83},
namely, we used the anticommutativity property of $\gamma_5$ only in
the box-type diagram and contract the string of $\gamma$-matrices
through $\gamma_5$, while in all other cases we did it in the other
direction such that no commutation with the chiral matrix occurred.

We have calculated the one-loop Feynman graphs represented in Fig.
\ref{NLO-NNA} in a $d = 4 - 2\epsilon$ dimensional space with the
following effective gluon propagator in the Landau gauge ($\mu^{*2}
= 4\pi \mu^2$):
\begin{equation}
(-i)\, \D^{ab}_{\mu\nu}(k)
=\frac{-i \delta^{ab}}{k^2 + i0}
\left(
\frac{\mu^{*2}}{- k^2}
\right)^\sigma
\left(
g_{\mu\nu} - \frac{k_{\mu}k_{\nu}}{k^2}
\right).
\end{equation}

A simple calculation leads to the results (since the sum of all
diagrams forms a gauge invariant set, only the $g_{\mu\nu}$-part
of propagator has been used in the real calculation)
\begin{eqnarray*}
T_{(1,a)}^A
&=&
f (\sigma, \epsilon)
\frac{1 - \epsilon}{(1 - x \omega)^{\sigma + \epsilon}}, \\
T_{(1,b)}^A
&=&
f (\sigma, \epsilon)
\frac{1}{(1 - \zeta \omega)^{\sigma + \epsilon}}
\biggl\{
\frac{1 + \epsilon (1 - \epsilon) - \epsilon
(\sigma + \epsilon)}{1 - \epsilon}
\left(
\frac{1 - \zeta \omega}{1 - x \omega}
\right)^{\sigma + \epsilon} \nonumber\\
&&- \frac{1}{1 + \sigma}
\left[
2 - \epsilon + 2\epsilon (\sigma + \epsilon)
- \frac{\sigma \epsilon (\sigma + \epsilon)}{1 - \epsilon}
\right]
{_2F_1} \left( \left.
{
1 + \sigma + \epsilon,\ 1 + \sigma
\atop
2 + \sigma
}
\right| \frac{( x - \zeta )\omega}{1 - \zeta \omega} \right)
\biggr\}, \\
T_{(1,c)}^A &=& T_{(1,b)}^A ( \zeta = 0), \\
T_{(1,d)}^A
&=&
f (\sigma, \epsilon)
(1 - x \omega) (1 - \epsilon)
\frac{\sigma ( 1 - \sigma) + 2 - 2\epsilon (1 - \epsilon) }{
(1 + \sigma) (2 + \sigma)}
F_1 \left( \left.
{
1 + \sigma + \epsilon,\ 1 + \sigma,\ 1
\atop
3 + \sigma
}
\right| x \omega, \zeta \omega \right),
\end{eqnarray*}
with the common overall factor $f(\sigma, \epsilon)$:
\begin{equation}
\label{f-function}
f (\sigma, \epsilon) =
- \frac{\alpha_s}{2 \pi} C_F
\left(
\frac{\mu^{*2}}{- Q^2}
\right)^{\sigma + \epsilon}
\frac{\Gamma (\sigma + \epsilon) \Gamma (2 - \epsilon) \Gamma
(1 - \sigma - \epsilon)}{ \Gamma (1 + \sigma) \Gamma (3 - \sigma
- 2 \epsilon)}
T_{(0)} ( \omega, x ) .
\end{equation}

While the calculation of the self-energy and vertex-type corrections
is straightforward we should mention some technical details about
the box-type graph. As quoted above a straightforward calculation
of the latter leads to a result in terms of the Appel function
$F_1$ \cite{BE53_1,PBM}. However, the latter can be reduced to the
difference of two hypergeometric functions ${_2F_1}$. The derivation
proceeds in a very simple manner. One evaluates the momentum integral
in $d$-dimensions with the integrand given by the product of the
denominators of the box diagram and the factor $(pk)$. There are two
possibilities to compute this integral: on the one hand we can join
all four denominators via Feynman parameters and get at the end the
result in terms of $F_1$. The second way is to represent the factor
$(kp)$ as a difference of two propagators, namely
\begin{equation}
(kp) = \frac{1}{2\zeta}
\left\{
[ ( x - \zeta )p - k ]^2 - [ xp - k ]^2
\right\}.
\end{equation}
Then, the final result looks like a difference of the vertex-type
graphs. Thus, we end up with the following relation between the
Appel function and hypergeometric ones which we have failed to
find in any textbook on special functions \cite{BE53_1,PBM}:
\begin{eqnarray}
&&F_1 \left( \left.
{
1 + a + b,\ 1 + a,\ 1
\atop
3 + a
}
\right| x, y \right) \\
&&\quad =
\frac{1}{y}\frac{( 2 + a )}{( a + b )}
\left\{
( 1 - y )^{-( a + b )}
{_2F_1} \left( \left.
{
a + b,\ 1 + a
\atop
2 + a
}
\right| \frac{x - y}{1 - y} \right)
-
{_2F_1} \left( \left.
{
a + b,\ 1 + a
\atop
2 + a
}
\right| x \right)
\right\}\nonumber\\
&&\quad =
\frac{1}{y}\frac{( 2 + a )}{( 1 - b )}
\left\{
{_2F_1} \left( \left.
{
1 + a + b,\ 1 + a
\atop
2 + a
}
\right| x \right)
-
( 1 - y )^{-( a + b )}
{_2F_1} \left( \left.
{
1 + a + b,\ 1 + a
\atop
2 + a
}
\right| \frac{x - y}{1 - y} \right)
\right\}.\nonumber
\end{eqnarray}

We put $\sigma = n \epsilon$, with $n$ being the number of the fermion
bubbles inserted in the gluon line, and multiply the above expressions
with a factor corresponding to the product of the
fermion vacuum polarization blobs
\begin{equation}
\pi_n = \frac{1}{\epsilon^n}
\left(
- \frac{\alpha_s}{4\pi}\beta_0
\right)^n
\left[ 6 \frac{ \Gamma ( 1 + \epsilon ) \Gamma^2 ( 2 - \epsilon )}{
\Gamma ( 4 - 2 \epsilon ) }
\right]^n.
\end{equation}
Performing trivial subtraction of sub- and overall divergences
and resummation, which is particularly easy as only the end terms
in the sum survive \cite{M-PPasqual}, along the line of
Ref. \cite{Braun94} we obtain the result given in the main text in
Eq. (\ref{FermionBubbleAmplitude}) with the functions $D_i$
corresponding to the contributions of particular diagrams
\begin{eqnarray}
D^A_{a} (\epsilon,\tau)
&=& T_{(0)} ( \omega, x )
\frac{ 1 - \epsilon }{( 1 - x \omega )^\tau} \\
D^A_{b} (\epsilon,\tau)
&=&
T_{(0)} ( \omega, x )
\frac{1}{(1 - \zeta \omega)^\tau}
\biggl\{
\frac{ 1 + \epsilon (1 - \epsilon) - \epsilon \tau}{1 - \epsilon}
\left(
\frac{1 - \zeta \omega}{1 - x \omega}
\right)^\tau \\
&&-
\frac{1}{1 - \epsilon + \tau}
\left[
2 - \epsilon + 2 \epsilon\tau
- \frac{\epsilon\tau (\tau - \epsilon)}{1 - \epsilon}
\right]
{_2F_1} \left( \left.
{
1 + \tau,\ 1 - \epsilon + \tau
\atop
2 - \epsilon + \tau
}
\right| \frac{(x - \zeta)\omega}{1 - \zeta \omega} \right)
\biggr\},\nonumber\\
D^A_{c} (\epsilon,\tau)
&=&
D^A_{b} (\epsilon,\tau| \zeta = 0), \\
D^A_{d} (\epsilon,\tau)
&=& T_{(0)} ( \omega, x )
\frac{ 1 - x \omega }{ \zeta\omega }
\left[
2 - \epsilon - \tau  + \frac{2 \epsilon \tau}{1 - \epsilon + \tau}
\right] \\
&\times&\left\{
{_2F_1} \left( \left.
{
1 + \tau,\ 1 - \epsilon + \tau
\atop
2 - \epsilon + \tau
}
\right| x \omega \right)
-
( 1 - \zeta \omega )^{- \tau}
{_2F_1} \left( \left.
{
1 + \tau,\ 1 - \epsilon + \tau
\atop
2 - \epsilon + \tau
}
\right| \frac{( x - \zeta )\omega}{ 1 - \zeta \omega } \right)
\right\}.\nonumber
\end{eqnarray}

Although our expression for the box-type diagram differs from the one
calculated (for $\zeta = 1$) in Ref. \cite{GosKiv97} by the factor
$\sim \epsilon\tau$ obviously this contribution has no impact, neither
on the final answer (\ref{FermionBubbleAmplitude}) since it vanishes
in both limits $\epsilon = \tau = 0$, nor on the one-loop results
derived in section \ref{PolarizedSector} as it proportional to
$\epsilon^2$.

%%%%%%%%%%%%%%%%%%%%%%%%%%%%%%%%%%%%%%%%%%%%%%%%%%%%%%%%%%%%%%%%%%%%%
\subsection{Spin-averaged scattering.}
%%%%%%%%%%%%%%%%%%%%%%%%%%%%%%%%%%%%%%%%%%%%%%%%%%%%%%%%%%%%%%%%%%%%%

For the spin-averaged amplitude there is no difficulty due to $\gamma_5$
and the calculation is straightforward
\begin{eqnarray*}
T_{(1,a)}^A &=& T_{(1,a)}^V, \qquad T_{(1,b)}^A = T_{(1,b)}^V,
\qquad T_{(1,c)}^A = T_{(1,c)}^V, \\
T_{(1,d)}^A
&=&
f (\sigma, \epsilon)
(1 - x \omega) (1 - \epsilon)
\frac{( 1 - \epsilon)( 2 - \epsilon ) -
( \sigma + \epsilon )(1 - \sigma - 3 \epsilon ) }{
(1 + \sigma) (2 + \sigma)} \nonumber\\
&&\hspace{5cm}\times F_1 \left( \left.
{
1 + \sigma + \epsilon,\ 1 + \sigma,\ 1
\atop
3 + \sigma
}
\right| x \omega, \zeta \omega \right),
\end{eqnarray*}
where the function $f(\sigma,\epsilon)$ is defined by
Eq. (\ref{f-function}).
Some simple manipulations give finally
\begin{eqnarray}
D^A_{a} (\epsilon,\tau) &=& D^V_{a} (\epsilon,\tau), \qquad
D^A_{b} (\epsilon,\tau) = D^V_{b} (\epsilon,\tau), \qquad
D^A_{c} (\epsilon,\tau) = D^V_{c} (\epsilon,\tau), \\
D^V_{d} (\epsilon,\tau)
&=& T_{(0)} ( \omega, x )
\frac{ 1 - x \omega }{ \zeta\omega }
\left[
2 - \epsilon + \tau  - \frac{4 \tau (1 - \epsilon)}{1 - \epsilon + \tau}
\right] \\
&\times&\left\{
{_2F_1} \left( \left.
{
1 + \tau,\ 1 - \epsilon + \tau
\atop
2 - \epsilon + \tau
}
\right| x \omega \right)
-
( 1 - \zeta \omega )^{- \tau}
{_2F_1} \left( \left.
{
1 + \tau,\ 1 - \epsilon + \tau
\atop
2 - \epsilon + \tau
}
\right| \frac{( x - \zeta )\omega}{ 1 - \zeta \omega } \right)
\right\}.\nonumber
\end{eqnarray}


\begin{thebibliography}{99}
\bibitem{Ji97}
X. Ji, Phys. Rev. Lett. 78 (1997) 610; Phys. Rev. D 55 (1997) 7114.
\bibitem{Rad96}
A.V. Radyushkin, Phys. Lett. B 380 (1996) 417; Phys. Lett. B 385
(1996) 333;
\bibitem{Rad97}
A.V. Radyushkin, Phys. Rev. D 56 (1997) 5524.
\bibitem{MRDGH}
D. M\"uller, D. Robaschik, B. Geyer, F.M. Dittes, J. Ho\v rej\v si,
Fortschr. Physik. 42 (1994) 101.
\bibitem{CFS96}
J.C. Collins, L.L. Frankfurt, M. Strikman, Phys. Rev. D 56 (1997) 2982.
\bibitem{renormalons}
For reviews and classic references see:\\
V.I. Zakharov, Nucl. Phys. B 385 (1992) 452;\\
A.H. Mueller, In {\it QCD 20 Years Later}, V. 1, World Scientific,
Singapore (1993);\\
The achievements in the field and the most recent references can be
found in:\\
B. Webber, {\it Renormalon phenomena in jets and hard processes},
hep-ph/9712236.
\bibitem{Braun94}
M. Beneke, V.M. Braun, Nucl. Phys. B 426 (1994) 301; Phys. Lett. B 348
(1995) 513;\\
P. Ball, M. Beneke, V.M. Braun, Nucl. Phys. B 452 (1995) 563.
\bibitem{FGG73}
S. Ferrara, R. Gatto, A.F. Grillo, {\it Conformal
algebra in space-time and operator product expansion}, Springer
tracts in modern physics, v. 67 (1973), and references cited
therein.
\bibitem{BelMul97}
A.V. Belitsky, D. M{\"u}ller, {\it Predictions from conformal
algebra for the deeply virtual Compton scattering}, Phys. Lett. B
(1998) (in press), hep-ph/9709379.
\bibitem{Jaffe83}
R.L. Jaffe, Nucl. Phys. B 229 (1983) 205.
\bibitem{DG98}
M. Diehl, T. Gousset, {\it Time ordering in off-diagonal parton
distributions}, hep-ph/9801233.
\bibitem{Rad83}
A.V. Radyushkin, Phys. Lett. B 131 (1983) 179.
\bibitem{EFP83}
R.K. Ellis, W. Furmanski, R. Petronzio, Nucl. Phys. B 212 (1983) 29.
\bibitem{Bel97}
A.V. Belitsky, B. Geyer, D. M{\"u}ller, A. Sch{\"a}fer, {\it On the
leading logarithmic evolution of the off-forward distributions}, Phys.
Lett. B (1998) (in press), hep-ph/9710427.
\bibitem{BGR87}
T. Braunschweig, B. Geyer, D. Robaschik, Ann. Phys. (Leipzig) 44 (1987)
403.
\bibitem{BB89}
I.I. Balitsky, V.M. Braun, Nucl. Phys. B 311 (1989) 541.
\bibitem{FFGS97}
L. Frankfurt, A. Freund, V. Guzey, M. Strikman, {\it Nondiagonal
parton distribution in the leading logarithmic approximation},
hep-ph/9703449.
\bibitem{kern}
J. Bl\"umlein, B. Geyer, D. Robaschik, Phys. Lett. B 406 (1997) 161.
\bibitem{BR97}
I.I. Balitsky, A.V. Radyushkin, Phys. Lett. B 413 (1997) 114.
\bibitem{Mank97}
L. Mankiewicz, G. Piller, T. Weigl, {\it Hard exclusive meson
production and non-forward parton distributions}, hep-ph/9711227.
\bibitem{DGPR97}
M. Diehl, T. Gousset, B. Pire, J.P. Ralston, Phys. Lett. B 411 (1997)
193.
\bibitem{JiOs97}
X. Ji, J. Osborne, {\it One-loop QCD corrections to the deeply-virtual
Compton scattering: the parton helicity-independent case},
hep-ph/9707254.
\bibitem{Mank97NLO}
L. Mankiewicz, G. Piller, E. Stein, M. V{\"a}nttinen, T. Weigl,
{\it NLO corrections to deeply virtual Compton scattering},
hep-ph/9712251.
\bibitem{BG95}
D.J. Broadhurst, A.G. Grozin, Phys. Rev. D 52 (1995) 4082.
\bibitem{IR_DIS97}
Yu.L Dokshitser, G. Marchesini, and B.R. Webber, Nucl. Phys. B469
(1996) 93; \\
E. Stein, M. Meyer-Hermann, L. Mankiewicz, A. Sch\"afer, Phys. Lett. B 376
(1996) 177; \\
M. Meyer-Hermann, M. Maul, L. Mankiewicz, E. Stein, A. Sch\"afer,
Phys. Lett. B 383 (1996) 463, (E) ibid. B 393 (1997) 487;\\
M. Maul, E. Stein, A. Sch\"afer, L. Mankiewicz, Phys. Lett. B 401
(1997) 100.
\bibitem{BE53_2}
H. Bateman, A. Erd\'elyi, {\it Higher transcendental functions}, V. 2,
Mc Graw-Hill, New York (1953).
\bibitem{BE53_1}
H. Bateman, A. Erd\'elyi, {\it Higher transcendental functions}, V. 1,
Mc Graw-Hill, New York (1953).
\bibitem{PBM}
A.P. Prudnikov, Yu.A. Brychkov, O.I. Marichev, {\it Integrals and Series:
Additional Chapters}, V. 3, Science, Moscow (1986).
\bibitem{Mue94}
D. M\"uller, Z. Phys. C 49 (1991) 293; Phys. Rev. D 49 (1994) 2525.
\bibitem{Mue97}
D. M\"uller, {\it Restricted conformal invariance in QCD and its
predictive power for virtual two-photon processes}, hep-ph/9704406.
\bibitem{CollinsBOOK}
J.C. Collins, {\it Renormalization}, Cambridge Univ. Press, Cambridge
(1984).
\bibitem{BM97}
A.V. Belitsky, E.A. Kuraev, Nucl. Phys. B 499 (1997) 301;\\
A.V. Belitsky, D. M\"uller, Nucl. Phys. B 503 (1997) 279;\\
A.V. Belitsky, {\it Leading order analysis of the twist-3 space-like
and time-like cut vertices in QCD}, XXXI PNPI Winter School
(1997), hep-ph/9703432.
\bibitem{AR88}
G. Altarelli, G.G. Ross, Phys. Lett. B 212 (1988) 391;\\
G.T. Bodwin, J. Qiu, Phys. Rev. D 41 (1990) 2755;\\
A.V. Manohar, Phys. Rev. Lett. 66 (1991) 289.
\bibitem{muel88}
F.M. Dittes, D. M\"uller, D. Robaschik, B. Geyer, Phys. Lett. B 209
(1988) 325.
\bibitem{ZVN94}
E.B. Zijlstra, W.L. van Neerven, Nucl. Phys. B 417 (1994) 61;
(E) ibid. B 426 (1994) 245; Nucl. Phys. B 383 (1992) 525;\\
W.A. Bardeen, A.J. Buras, D.W. Duke, T. Muta, Phys. Rev. D 18
(1978) 3998.
\bibitem{Yndurain79}
A. Gonzalez-Arroyo, C. Lopez, F.J. Yndurain, Nucl. Phys. B 153 (1979)
161;\\
A. Devoto, D.W. Duke, Riv. Nuovo Cim. 7 (No. 6) (1984) 1.
\bibitem{MikhsRad}
S.V. Mikhailov, A.V. Radyushkin, {\it Evolution kernel for the pion
wave function: two-loop QCD calculation in Feynman gauge}, JINR
preprint P2-83-271 (1983); Nucl. Phys. B 254 (1985) 89.
\bibitem{Gra97}
J.A. Gracey, Phys. Lett. B 322 (1994) 141; Nucl. Phys. B 480 (1996) 73.
\bibitem{Mankiewicz97}
L. Mankiewicz, M. Maul, E. Stein, Phys. Lett. B 404 (1997) 345.
\bibitem{Mik97}
S.V. Mikhailov, {\it Renormalon chains contributions to the non-singlet
evolution kernels in $[\varphi^3]_6$ and QCD}, hep-ph/9706326.
\bibitem{GosKiv97}
P. Gosdzinsky, N. Kivel, {\it Resummation of $(-b_0\alpha_s)^n$
corrections to the photon-meson transition form-factor $\gamma^*
\gamma \to \pi^0$ }, hep-ph/9707367.
\bibitem{Mue95}
D. M\"uller, Phys. Rev. D 51 (1995) 3855.
\bibitem{Bra83}
E. Braaten, Phys. Rev. D 28 (1983) 524;\\
E.P. Kadantseva, S.V. Mikhailov, A.V. Radyushkin, Sov. J. Nucl. Phys.
44 (1986) 326.
\bibitem{M-PPasqual}
A. Palanques-Mestre, P. Pascual, Comm. Math. Phys. 95 (1984) 277.
\bibitem{Lipatov}
L.N. Lipatov, Sov. Phys. JETP, 45 (1977) 216;\\
E.B. Bogomolny, V.A. Fateyev, Phys. Lett. B 71 (1977) 93.
\bibitem{Radyushkin-model}
A.V. Radyushkin, quoted in Ref. \cite{Mank97} and in preparation.
\bibitem{JiMelSon97}
X. Ji, W. Melnitchouk, X. Song, Phys. Rev. D 56 (1997) 5511.
\bibitem{Pol97}
V.Yu. Petrov, P.V. Pobylitsa, M.V. Polyakov, I. B\"ornig, K. Goeke,
C. Weiss, {\it Off-forward quark distributions of the nucleon in the
large $N_c$ limit}, hep-ph/9710270.
\bibitem{JiOs98}
X. Ji, J. Osborne, {\it One-loop corrections and all order
factorization in deeply virtual Compton scattering}, hep-ph/9801260.
\bibitem{Collins98}
J.C. Collins, A. Freund, {\it Proof of factorization for deeply virtual
Compton scattering in QCD}, hep-ph/9801262.
\end{thebibliography}
\end{document}